\documentclass[10pt,a4paper]{article}
\usepackage{jheppub}
\usepackage[utf8]{inputenc}
\usepackage[english]{babel}
\usepackage{amsmath}
\usepackage{amsfonts}
\usepackage{comment}
\usepackage{amssymb}
\usepackage{bm}
\usepackage{setspace}

\newcommand{\SOMMA}[2]{\displaystyle\sum\limits_{#1}^{#2}}
\newcommand{\bb}[1]{\boldsymbol{#1}}

\def\l{\langle}
\def\r{\rangle}

\title{Dense auto-hetero associative memories applied to noisy communication channels.}
\author[a,1,2,3]{Elena Agliari,}
\author[b,1,4]{Andrea Alessandrelli,}
\author[c,1,2,3]{Adriano Barra,}
\author[a,1]{Alberto Fachechi,}

\affiliation[a]{Dipartimento di Matematica, Sapienza Università di Roma, Italy}
\affiliation[b]{Dipartimento di Matematica e Fisica, Università del Salento, Italy}
\affiliation[c]{Dipartimento di Scienze di Base e Applicazioni all'Ingegneria, Sapienza Università di Roma, Italy}
\affiliation[1]{GNFM, Istituto Nazionale d'Alta Matematica, Sezione di Roma, Italy}
\affiliation[2]{INFN, Istituto Nazionale di Fisica Nucleare, Sezione di Roma, Italy}
\affiliation[3]{CNR, Nanotec, Salento Unit, Lecce, Italy}
\affiliation[4]{INFN, Istituto Nazionale di Fisica Nucleare, Sezione di Lecce, Italy}

\emailAdd{alberto.fachechi@uniroma1.it}
\abstract{
Networks of interacting Hebbian networks have recently been shown to perform a task beyond associative memory, namely \emph{pattern disentanglement}: when fed with a spurious mixture of stored patterns, the different modules spontaneously specialize on, and retrieve, the different constituents of the mixture. So far, this capability has only been established for pairwise interactions, which limits the number of patterns that can be handled. Here we introduce a dense extension of these modular networks, in which both the auto-associative couplings within each module and the hetero-associative couplings among modules are promoted to higher-order Hebbian interactions. We show that, with a suitable choice of the interaction orders, the network disentangles mixtures while storing a number of patterns that scales linearly with the module size, a regime where its pairwise counterpart fails. Through a statistical-mechanical analysis based on Guerra's interpolation, we derive the self-consistency equations for the order parameters and draw the phase diagrams identifying the region where disentanglement is achieved; these predictions are confirmed by Monte Carlo simulations. Finally, we show that disentanglement provides a natural decoding primitive, and we illustrate it with two applications: the explicit reconstruction of all the hidden patterns from the Hebbian tensors and a stream of unlabeled mixtures, and a proof-of-concept communication protocol in which each message token is transmitted as a masked mixture of hidden patterns and decoded by the network dynamics. Owing to its attractor-based decoding, the protocol degrades gracefully under strong channel corruption, where conventional secure-transmission pipelines fail abruptly.
}

\toccontinuoustrue

\begin{document}

\newpage

\maketitle

\section{Introduction and context}

Associative memories are among the most successful paradigms for understanding collective information processing in neural networks. Since the seminal work of Hopfield \cite{hopfield1982neural}, they have provided a statistically-mechanical framework for studying memory retrieval, error correction, and distributed computation, while simultaneously inspiring a large body of modern machine-learning architectures \cite{Lenka,Carleo,DeepLearners}. 

Over the last four decades, two major research directions have substantially extended the original framework:
\newline
The first is represented by \emph{dense} (or modern) Hopfield networks \cite{Pedreschi,GuerraDenso,Dima2,Dima3,Dima4,Dima1}, where the standard pairwise Hebbian interaction is generalized to higher-order couplings. Besides considerably enlarging the storage capacity of associative memories \cite{Dima1,ExpHop1,Lowe,ExpHop3,ExpHop4}, dense interactions have been shown to improve robustness against adversarial perturbations \cite{Dima2}, enhance the signal-to-noise ratio during retrieval \cite{agliari2020neural}, and provide a flexible computational framework for modern associative memories \cite{Lowe}.
\newline
The second development concerns \emph{hetero-associative} memories \cite{BAM_Kosko,Catania,cooperative_TAM,sup_TAM_Hard,ICLR_TAM}, in which neurons arranged in different layers, or modules, interact through associative couplings. Unlike Boltzmann machines, whose inter-layer connections are learned via optimization procedures, hetero-associative memories retain the Hebbian nature of their synapses and these are designed so that neurons in different modules retrieve distinct, associated patterns -- hence the name. As a result, their cooperative dynamics naturally enables generalized pattern-recognition, transfer of information across modules, and the processing of heterogeneous data representations.

\begin{figure}[t]
    \centering
    \includegraphics[width=0.9\textwidth]{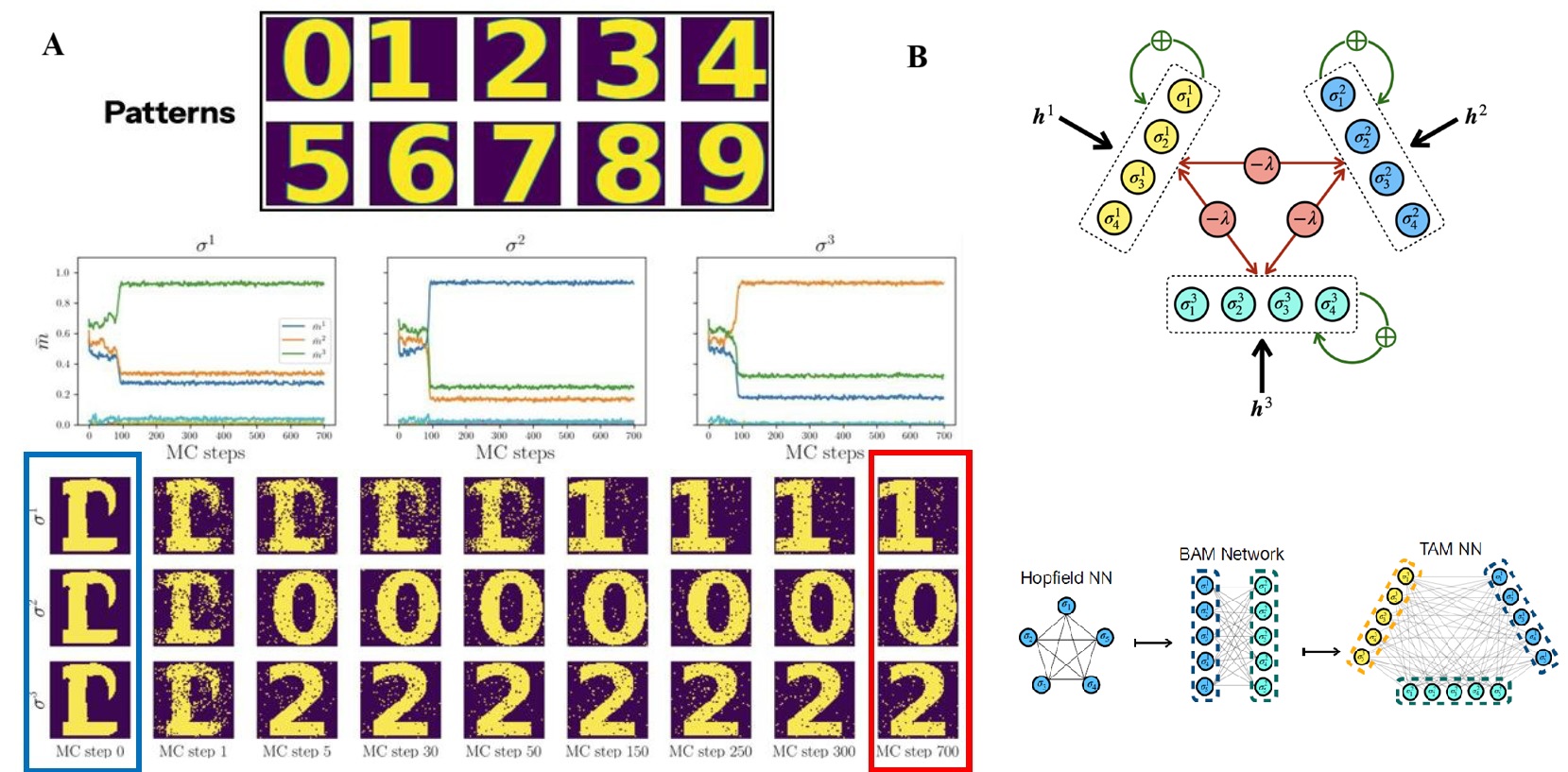}
    \caption{Cartoon summarizing the pattern disentanglement task by these generalized neural networks (taken from \cite{N_of_NN}. (A) The patterns to handle are the standard digits shown on top. The network is initialized in a 3-pattern mixture state, as highlighted by the the blue rectangle in the left bottom plot; these kind of mixtures are traditionally regarded as retrieval errors in Hopfield network, its Achilles heel. Next, this initial configuration is updated according to the dynamics \eqref{eq:update_MC_iclr} upon convergence to a final configuration where the mixture has been disentangled and, on each layer (one per layer) the patterns giving rise to the mixture are retrieved. (B) On top the three-module architecture used to accomplish this task, while, in the bottom line, the extension from the standard Hopfield network, toward the BAM (Bi-directional Associative Memory) and the TAM (Three-Directional Associative Memory) is shown and acts as the basis of the present $L$-directional auto-hetero-associative generalization.}
    \label{fig:mnist-example}
\end{figure}

In the present paper we merge these two directions. In particular, we will consider \emph{networks of associative memories} \cite{N_of_NN}, where several Hopfield modules interact through (repulsive) hetero-associative couplings while retaining (attractive) auto-associative interactions within each module: the competition between these two mechanisms yield a novel emergent computational capability, termed \emph{pattern disentanglement}: starting from a spurious mixture state, the different modules spontaneously specialize on, and thereby recover, the different constituent patterns hidden in the composite signal, see Fig.~\ref{fig:mnist-example}.
As long as neural interactions are pairwise, the affordable load is relatively low \cite{N_of_NN,sup_TAM_Hard}. Here, in order to handle a number of patterns scaling linearly with the number of neurons, we densify these modular networks: the auto-associative interactions within each module and the hetero-associative interactions among different modules are both promoted to higher-order Hebbian couplings. This choice is not only computationally advantageous, but also analytically convenient. Indeed, while a dense network with $P$-body interactions can store up to $K = \Theta(N^{P-1})$ patterns, as we restrict to the regime $K =\Theta(N)$, we can enjoy the benefits of a low-load regime (relative to the maximal capacity of the architecture) and, in particular, the statistical-mechanical analysis can be performed using simple order parameters, without resorting to the full machinery of spin-glass theory required at saturation \cite{Parisi1,Parisi2}.

Within this framework, we obtain a statistical mechanical picture of these dense modular auto/hetero-associative memories, identify the region in the space of the control parameters supporting pattern disentanglement, and validate the analytical predictions through numerical simulations.
Finally, motivated by the natural interpretation of disentanglement as a decoding operation, we illustrate two applications of the proposed architecture inspired by cryptography.

\section{The model and its related statistical mechanical description}
\label{sec:model_rs}

The global network is made up of $L$  interacting Hopfield networks  -- referred to as {\em modules} -- each made of $N$ binary neurons\footnote{Heterogeneity in the size of the modules could be accounted as a minor variation on theme but it does not alter substantially the picture we obtain and it will not be deepened in this paper.} such that the configuration of each module $a\in\{1,\dots,L\}$ is denoted by
$\bm{\sigma}^{a} = (\sigma^{a}_1,\dots,\sigma^{a}_{N})\in\{-1,+1\}^{N},
$ 
and the full network state is identified by the matrix $\bm{\sigma}=(\bm{\sigma}^1,\dots,\bm{\sigma}^L)\in\Omega$, with $\Omega=\{-1,+1\}^{L\times N}$. The set of stored patterns is denoted as $\{\bb\xi^{\mu}\}_{\mu=1}^K$ and consist of  $K$ vectors of length $N$, whose entries are Rademacher random variables. 

The retrieval of a given pattern $\mu$ by a given module $a$ is measured by the module-wise Mattis magnetizations
\begin{equation}
m_{\mu}^{a}(\bm{\sigma})
=\frac{1}{N}\sum_{i=1}^{N}\xi_i^\mu \sigma_i^{a},
\label{eq:Mattis_TAM}
\end{equation}
for all $ a=1,\dots,L$ and $ \mu=1,\dots,K$; in the present setting, these observables are the solely order parameters needed to describe the macroscopic behavior of the model. 
\newline
In order to build the cost function of this system, we point out that
\begin{itemize}
    \item  The dense auto-associative terms, each built of by interactions of order $P$, can be implemented in the cost function via terms $\propto \sum_{\mu,a}(m_\mu ^a)^P$, with $P>2$ (to avoid the pairwise limit) and  $P$ even (to preserve the symmetry $\bb\sigma^a\to -\bb\sigma^a$).
    \item The dense hetero-associativity terms, each built of by interactions of order $D$, can be accounted by including in the cost function contributions of the form $\propto \sum_{\mu, a \neq b}(m_\mu ^a m_\mu ^b)^{D/2}$: we require $D/2$ to be even in order for the model to be invariant under module-wise spin-flip symmetry (namely $\bb\sigma^a\to -\bb \sigma^a$ for a given $a=1,\dots,L$, with all other modules unchanged)\footnote{With this choice, configurations as $(\bb\xi^1,\bb\xi^1,\bb\xi^1)$ and $(\bb\xi^1,\bb\xi^1,-\bb\xi^1)$ share the same energy, with beneficial effects on disentangling: see  \cite{N_of_NN}.}. 
    \item The $n$-body Hebbian couplings are defined as 
    \begin{equation} 
    J^{(n)}_{i_1,\dots,i_n} =\frac{1}{N^{n-1}}\sum_{\mu=1}^{K}\xi_{i_1}^\mu\cdots\xi_{i_n}^\mu.
    \end{equation}
\end{itemize}
By merging these items, we introduce
the global cost function of the network $\mathcal{H}_{N,L}^{(P,D)}(\bm{\sigma}\,|\,\bm{h},\bm{J},\lambda,\theta)$ as
\footnote{In this definition, we do not adopt the standard normalization convention for dense associative memories, which would introduce factors of $1/P!$ and $1/D!$ in the auto- and hetero-associative contributions, respectively. Consequently, the effective scales of $\beta$ and $\lambda$ explicitly depend on the interaction orders. This does not affect our analysis, as our main goal is to establish that dense modular networks exhibit disentangling capabilities in the regime $K=\alpha N$.}
\begin{align} 
\mathcal{H}_{N,K,L}^{(P,D)}(\bm{\sigma}\,|\,\bm{h},\bm{J},\lambda,\theta)
=&\;
-\frac{1}{2}\sum_{a=1}^{L}\sum_{i_1,\dots,i_P=1}^{N}
J^{(P)}_{i_1,\dots,i_P}\,\sigma^{a}_{i_1}\cdots\sigma^{a}_{i_P}
\\
&\;
+\frac{\lambda}{2}\sum_{a=1}^{L}\sum_{b>a}^{L}\sum_{i_1,\dots,i_D=1}^{N}
J^{(D)}_{i_1,\dots,i_D}\,
\sigma^{a}_{i_1}\cdots\sigma^{a}_{i_{D/2}}\,
\sigma^{b}_{i_{D/2+1}}\cdots\sigma^{b}_{i_D}
\label{eq:hamiltonian_LAM}\\
&\;
- \theta\sum_{a=1}^{L}\sum_{i=1}^{N} h_i^{a}\sigma_i^{a},
\notag
\end{align}
where $J^{(P)}$ rules the auto-associative part while $J^{(D)}$ the hetero-associative counterpart; moreover, $\lambda\in\mathbb{R}^+$ rules the relative weight between the intra- and the inter-module interactions, and $\theta\in\mathbb{R}$ tunes the magnitude of the signal conveyed to the network in terms of  the external fields $\bm{h}^a=(h_1^a,\dots,h_N^a)\in\{-1,+1\}^{N}$. Note the {\em reverse sign} in front of $\lambda$: this way the hetero-association works as a penalty term disfavoring alignment to the same pattern by two distinct modules and, as a result, each module is encouraged to specialize on a different memory pattern, thus promoting disentanglement. In the following, in order to lighten the notation, we will drop the dependence on the structure parameters $(N, K, L, P, D)$, on the control parameters $(\lambda, \theta)$, and on the task-related variables $(\boldsymbol J, \boldsymbol h)$.

Before proceeding, we comment on the choice of the interaction orders. The storage capacity of the network is ultimately determined by the lowest interaction-order, consequently, setting only the auto-associative couplings to order higher than two, while keeping the hetero-associative ones pairwise, would provide little benefit, as the latter would still constitute the limiting factor. This motivates the choice of making both auto- and hetero-associative interactions dense.
\newline
Furthermore, two distinct sources of slow noise affect the retrieval dynamics now. In standard (single-module) auto-associative memories, slow noise originates from the correlations induced by the Hebbian superposition of many stored patterns. In the present architecture, instead, an additional contribution arises from the interaction among different modules through the hetero-associative couplings. 
Remarkably, by choosing appropriately  $P$ and $D$ we can suppress both the effects while preserving a linearly extensive storage regime, $K = \Theta(N)$. In fact, setting $P>2$ corresponds to a low-load regime where pattern-induced slow noise becomes negligible; if we also require $P \leq D$, namely that hetero-associative interactions are at least as dense as the auto-associative ones, the module-induced noise is suppressed even more strongly than the pattern-induced contribution. As a result, setting $2 < P \leq D$ allows us to deal with a linearly extensive number of patterns still remaining within a simple regime. However, nothing comes for free: the drawback is that a larger amount of information needs to be stored as $J^{(P)}$ and $J^{(D)}$ overall encompass $\mathcal O(N^P)$ and $\mathcal O(N^D)$ entries. Here these entries are deterministically determined by Hebb's rule, yet in a machine-learning context their determination by a training protocol could be challenging.
\par\medskip
Exploiting the mean-field structure of the model, the cost function \eqref{eq:hamiltonian_LAM} can be conveniently recast as $\mathcal{H}(\bm{\sigma})=
- \sum_{a=1}^{L}\sum_{i=1}^{N} \hat{h}_i^{a}\sigma_i^{a} $
with the effective field $\hat {\bb h}^{a}$ acting on neurons in the module $a$ being the sum of three contributions:
\begin{equation}
\label{eq:net_field}
\hat{\bm h}^{a}(\bb\sigma)
=
\bm h^{a\to a}(\bb\sigma)
+
\sum_{\substack{b=1 \\ b\neq a}}^L\bm h^{b\to a}(\bb\sigma)
+ 
\theta \,\bm h^{a},
\end{equation}
respectively, the intra-module (auto-associative), inter-module (anti-imitative) and external fields. The definitions of the first two contributions stem directly from the expression (\ref{eq:hamiltonian_LAM}), that is
\begin{eqnarray}
    h_{i_1}^{a\to a}(\bb\sigma) &=& \frac{P}{2}\sum_{\underset{(i_2, \cdots, i_P)\neq i_1}{i_2,\dots,i_P=1}}^{N}
J^{(P)}_{i_1,\dots,i_P}\,\sigma^{a}_{i_2}\cdots\sigma^{a}_{i_P},\label{eq:intra}\\
    h_{i_1}^{b\to a}(\bb\sigma)&=&
-\frac{\lambda D}{2}\sum_{\underset{(i_2, \cdots, i_D)\neq i_1}{i_2,\dots,i_D=1}}^{N}
J^{(D)}_{i_1,\dots,i_D}\,
\sigma^{a}_{i_2}\cdots\sigma^{a}_{i_{D/2}}\,
\sigma^{b}_{i_{D/2+1}}\cdots\sigma^{b}_{i_D},\label{eq:inter}
\end{eqnarray}
while the external field $\boldsymbol h^a$ can be used to encode the task-dependent cue, as discussed in the next subsection.

We can now use the cost function \eqref{eq:hamiltonian_LAM} to generate a Markov chain for the neural dynamics: introducing the inverse temperature $\beta=T^{-1}\in\mathbb{R}^+$ to tune the fast noise in the network, the stochastic process governing the neural dynamics reads as (e.g., see \cite{CKS})
\begin{equation}
\label{eq:update_MC_iclr}
\bm\sigma^{a}(t{+}1)
=
\operatorname{sign}[\tanh(\beta\hat{\bm h}^{a}(\bb\sigma(t))) + \bm u^{a}(t)],
\end{equation}
where $\hat{\bb h}^a$ is the $a$-th module effective field, computed at each time-step $t$ according to Eqs. (\ref{eq:net_field}-\ref{eq:inter}), and 
$\bm u^{a}(t)\underset{i.i.d}{\sim} \mathcal{U}([-1,1]^N)$ for all $a=1,\dots,L$. The dynamics in Eq. \eqref{eq:update_MC_iclr} will be used in Monte Carlo simulations, see Sec. \ref{sec:numerics}.

Our goal now is to characterize the macroscopic behavior of the system by analyzing the evolution of its order parameters in the space of the control parameters, ultimately resulting  in a {\em  phase diagram}, possibly revealing the existence of a non-empty region in which the pattern disentanglement task can be reliably accomplished. To extract this knowledge from the cost function we follow a statistical-mechanics approach (see Sec.~\ref{subsec:free_energy_rs}): i) first we obtain an explicit expression of the free-energy associated to the cost function in terms of its control and order parameters; ii) its extremization w.r.t. the order parameters then returns a set of self-consistency equations (i.e., a set of input-output relations ruling the network's behavior), whose solution captures how the network responds to different external stimuli and also how robust such responses are as the control parameters are tuned. Once the analytical journey is over, Monte Carlo simulations are run independently to confirm the obtained picture (see Sec.~\ref{sec:numerics}).

\subsection{Statistical-mechanical analysis: scaling regimes and RS solution}
\label{subsec:free_energy_rs}

Before presenting the statistical-mechanics analysis it is worth recalling the control parameters of the model, namely $\beta$ (controlling the noise in the neuronal dynamics), $\theta$ and $\lambda$ (tuning the magnitude of the contributions appearing in the cost function); further, as we are dealing with a linearly extensive number of patterns  $K =\Theta(N)$, it is convenient to define $\alpha \doteq \lim_{N\to\infty}\frac{K}{N} > 0$ for fine-tuning the load (this will be helpful in the finite-size simulations, while in the theoretical investigation carried on in the thermodynamic limit, this parameter will not be relevant). We also recall the three structural parameters, namely the  densities $P$ and $D$ of the, respectively, auto- and hetero- associative couplings, and the number of modules $L$\footnote{Generalizing this minimal setting toward more heterogeneous scenarios is straightforward (for instance, the density of the auto-associative couplings $P$ can be made module dependent $P(a)$ and/or the density of the hetero-associative couplings $D$ among neurons in moduli $a,b$ can be made moduli dependent $D(a,b)$), but we will not deepen these variations on theme in this first paper.}. 
\par\medskip
{\em En route} for a statistical mechanical analysis, we endow the configuration space $\Omega$ of the model with the (random) Boltzmann-Gibbs measure
\begin{eqnarray}
    \mathcal{P}(\bm{\sigma}; \beta)
&=&\frac{\exp\big[-\beta\,\mathcal{H}(\bm{\sigma})\big]}
{\mathcal{Z} (\beta)},
\label{BGmeasure}\\
\mathcal{Z}(\beta)
&=&\sum_{\bm{\sigma}\in\Omega}
\exp\big[-\beta\,\mathcal{H}(\bm{\sigma})\big].
\label{partition-function_TAM_easy}
\end{eqnarray}
%
The associated quenched free energy in the thermodynamic limit is defined as
\begin{equation}
\mathcal{A}(\beta)
=\lim_{N\to\infty}
\frac{1}{N}\mathbb{E}_{\bm{\xi}}\ln \mathcal{Z}(\beta),
\label{eq:Free-Definition}
\end{equation}
with $\mathbb E_{\bb\xi}$ being the expectation value w.r.t. the pattern realization. To inspect the disentangling task we need to specify the signal $\boldsymbol h$ conveyed to the model and, accordingly, the neuronal configurations fulfilling the task and whose equilibrium must be studied. The signal is taken as a symmetric mixture of $L$ distinct patterns, namely, without loss of generality, those with label $\mu=1, ..., L$. Thus, we focus on the class of states displaying a Mattis magnetization $ {\bb m}^a = ( m_1 ^a, \dots,  m_L^a,0,\dots,0)$. Within this setting, we assume that the order parameters self-average around their means in the thermodynamic limit (as expected given that we are operating in the low-load regime) and we adapt Guerra's interpolation to get the explicit expression of the quenched free energy  in terms of the control and order parameters (as detailed in App.~\ref{app:guerra_interpol}) 
\begin{equation}
\label{eq:final_stat_pressure_RS_MAIN}
\begin{array}{lll}
         \mathcal{A}(\beta) &=& \SOMMA{a=1}{L}\mathbb{E}_{\bm\xi}\log 2 \cosh\Bigg[\beta\SOMMA{\mu=1}{L}\Big( \dfrac{P}{2}(\bar{m}^a_\mu)^{P-1}  - \sum\limits_{b\neq a}^L\lambda\dfrac{D}{2}(\bar{m}^a_\mu)^{D/2-1}(\bar{m}^b_\mu)^{D/2}\Big)\xi^\mu + \beta \theta h^a\Bigg]
         \\\\
         &-&\dfrac{\beta}{2}(P-1)\SOMMA{\mu=1}{L}\SOMMA{a=1}{L}\Bigg[(
         \bar{m}^a_\mu)^P -\lambda\dfrac{D-1}{P-1}\SOMMA{b\neq a}{L}(\bar{m}^a_\mu \bar{m}^b_\mu\big)^{D/2} \Bigg],
\end{array}
\end{equation}
where $\bar m_\mu ^a = \lim_{N\to\infty} \mathbb E_{\bb\xi} \omega(m_\mu ^a)$ and $\omega(\cdot)$ is the Boltzmann-Gibbs expectation associated with the probability measure introduced in Eqs. (\ref{BGmeasure}-\ref{partition-function_TAM_easy}).
\newline
Extremizing the free energy with respect to the order parameters yields a closed set of self-consistent equations ruling their evolution in the control-parameter space
\begin{equation}
\bar{m}^a_\nu =\mathbb{E}_{\bm\xi}\xi^\nu\tanh\Big[\beta\SOMMA{\mu=1}{L}\big( \dfrac{P}{2}(\bar{m}^a_\mu)^{P-1}  - \sum\limits_{b\neq a}^L\lambda\dfrac{D}{2}(\bar{m}^a_\mu)^{D/2-1}(\bar{m}^b_\mu)^{D/2}\big)\xi^\mu  +\beta \theta h^a\Big],
\label{eq:self_TAM_dense}
\end{equation}
where $h^a$ denotes a generic field entry for module $a$. 
Note that these equations can be seen as {\em input-output} relation, namely, once specified the structure of the signal provided to the network (i.e., the input), they return the global response of the network (i.e., the output). Also note that, since the model operates in its low-storage regime, these equations do not depend on the load $\alpha$.

\begin{figure}[h]
\centering
\includegraphics[width=0.45\textwidth]{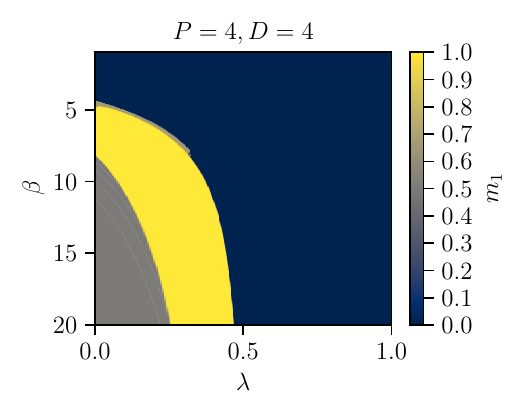}
\includegraphics[width=0.45\textwidth]{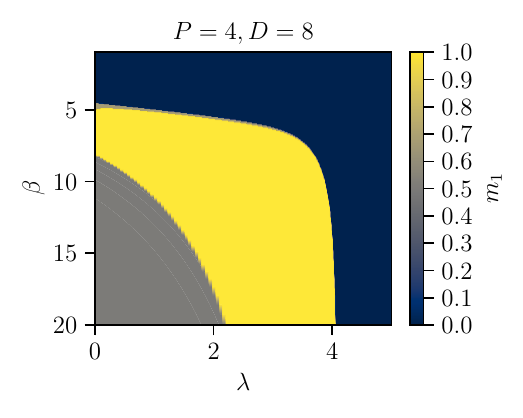}
\caption{{\bfseries Phase diagram for the disentangling task}. We report the equilibrium value of the magnetization $\bar{m}_1$, obtained by solving the self-consistency equations \eqref{eq:self_TAM_dense} and adopting the state $\bar{m}^{{mix}}$ as the starting point of the fixed-point iteration method. We set $L=3$, $\theta=3$, and $P=D=4$ (left) or $P=4$, $D=8$ (right). Three different scenarios emerge as the parameters $(\beta, \lambda)$ are tuned: an ergodic phase ($\bar{m}_1 \approx 0$, blue region), a spurious phase ($\bar{m}_1 \approx 0.5$, gray region) in which the system remains in the state $\bar{m}^{\mathrm{mix}}$, and a disentangled region ($\bar{m}_1 \approx 1$, yellow region) in which the system moves from $\bar{m}^{{mix}}$ to $\bar{m}^{{dis}}$.}
\label{fig:TAM-phase-diag}
\end{figure}

Recalling the previous ansatz on the structure of the solution, the self-consistency equations can be organized in a $L \times L$ matrix $\bar {\bb m}$ whose $(\mu,a)$-entry corresponds to the Mattis magnetization w.r.t. the pattern with index $\mu \le L$ at module $a\le L$. In particular, the disentangled state would correspond to a diagonal form of $\bar {\bb m}$, as each given module specializes on a different pattern. Conversely, the symmetric spurious state of $L$ patterns is characterized by a matrix with all non-vanishing entries.
As an example, for $L=3$ these two solutions correspond, respectively, to
$$
\bar {\bb m}^{dis}=\begin{pmatrix}
\bar m &0 &0\\
0 &\bar m &0\\
0 &0 &\bar m
\end{pmatrix},\quad \quad\quad
\bar {\bb m}^{mix}=\begin{pmatrix}
\bar m &\bar m &\bar m\\
\bar m &\bar m &\bar m\\
\bar m &\bar m &\bar m
\end{pmatrix},
$$
where the symmetry in the input is mirrored by homogeneous entries.
Then, the value of $\bar m$ is obtained by plugging $\bar {\bb m}^{dis,mix}$ in the self-consistency equations: solving \eqref{eq:self_TAM_dense} determines its  behavior  as a function of the control parameters $(\beta,\lambda,\theta)$ as well as the structural parameters $(P,D,L)$ and allows us to draw the phase diagram: its projection on the most relevant control parameters for the disentangling task  (i.e., on the $\beta, \lambda$ plane) is shown in Fig.~\ref{fig:TAM-phase-diag}. Specifically, the disentangling phase (i.e., the yellow region) is identified as the intersection of the region where spurious mixtures are unstable states and the region where disentangled configurations are stable states: for further details about the realization of these diagrams, see App. \ref{app:guerra_interpol}.


\subsection{Numerical comparison: disentangling in the linearly extensive regime $K = \Theta(N)$}
\label{sec:numerics}

As the phase diagram is available, we can force the network to operate within its disentangling region (i.e., the yellow areas of the plots reported in Fig.~\ref{fig:TAM-phase-diag}) and inspect its performances: this way we can appreciate by numerical evidence how the dense model outperforms its pairwise counterpart and check the robustness of theoretical results versus finite-size effects.
\newline
In particular, we consider incoming fields that are mixtures of $L$ patterns (again, without loss of generality, we are focusing on the first $L$ patterns\footnote{In principle, the number of mixture components $M$ need not coincide with the number of layers. If $M>L$, the neural network is expected to reproduce a subset of patterns involved in the spurious combinations. In this case, due to the stochasticity of neural dynamics at finite $\beta$ one can repeatedly perform thermalization to reproduce all the patterns involved in the mixture states (and discard possible duplicates by checking the mutual overlap between reconstructed patterns). Conversely, if $M<L$ and depending on the value of $\beta$ and $\lambda$, some modules could align on the same pattern or retrieve memories not involved in the combination. In the latter case, one can reject these extra components by checking the overlap between them and the starting initial condition. Notice that one can always estimate $M$ by checking the overlap with the stored patterns (if they are known) or their energy value as in the pairwise case \cite{ICLR_TAM,bumi}.})
\begin{equation}
\boldsymbol h^a =  \textrm{sign}(\sum_{\mu =1}^L \boldsymbol \xi^{\mu}), ~\mathrm{for}~ a=1,...,L,
\end{equation}
and follow the evolution of the neuronal configuration $\boldsymbol \sigma$ updated according to the neural dynamics \eqref{eq:update_MC_iclr}. Given the task that we have in mind, it is natural to take as the initial conditions for the neural dynamics
\begin{equation} \label{eq:init}
\boldsymbol \sigma^{a}(t=0)=   \textrm{sign}(\sum_{\mu =1}^L \boldsymbol \xi^{\mu}), ~\mathrm{for}~ a=1,...,L.
\end{equation}
The neural dynamics \eqref{eq:update_MC_iclr}, initialized according to \eqref{eq:init}, can spontaneously converge to neural configurations where different modules align with the different patterns composing the mixture, provided that we confine this problem in the linear setting $K = \alpha N$ and we properly choose $\lambda$ and $\beta$. 

\begin{figure}
    \centering
    \includegraphics[width=\linewidth]{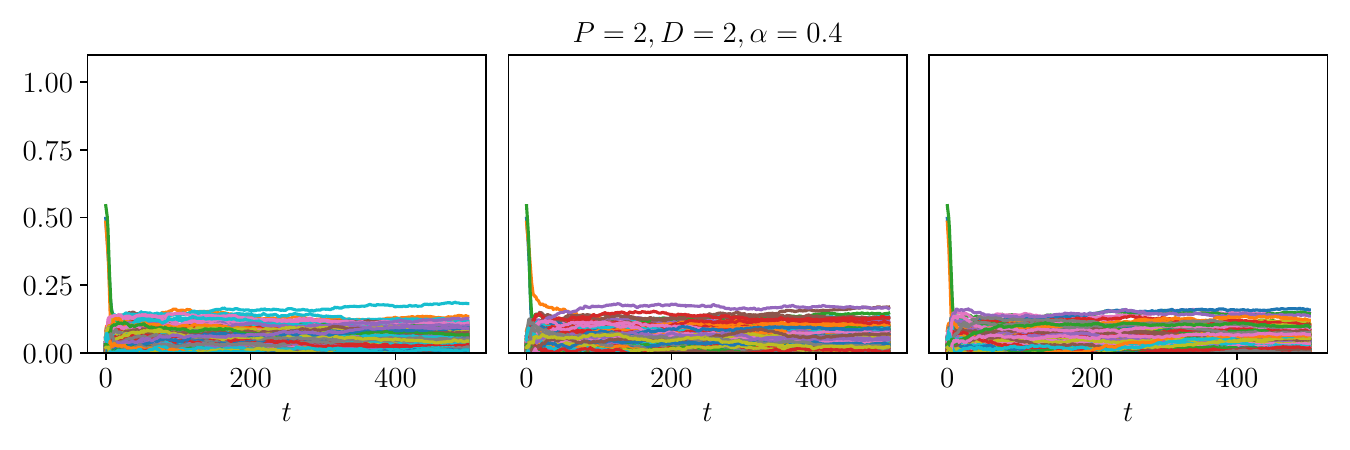}
    \includegraphics[width=\linewidth]{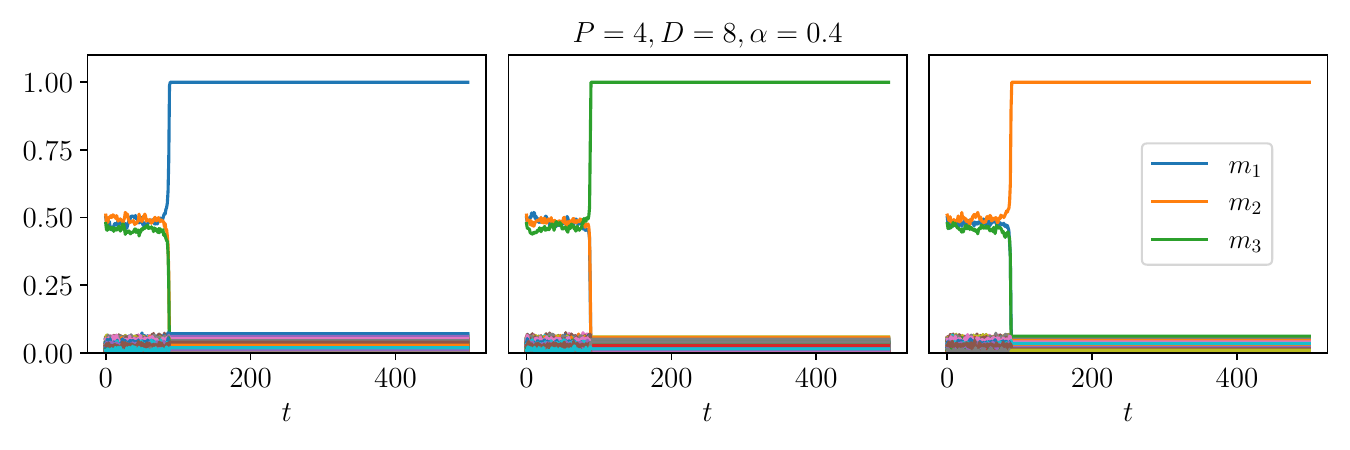}
    \caption{{\bfseries Examples of disentangling dynamics.} The two series of plots refer to a disentanglement experiment for the pairwise limit $P=D=2$   (first row) and the present dense extension $P=4$ and $D=8$ (second row) of this network of Hebbian networks. In all cases, $N=2000$ and $K=\alpha N$ with $\alpha=0.4$, $L=3$ and the initial condition is a 3-mixture spurious state ${\bb\sigma_3} = \operatorname{sign}({\bb\xi^1+\bb\xi^2+\bb\xi^3})$. Each column refers to the magnetizations as a function of evolution time w.r.t. all the patterns of each module in the model. The external field is $\bb h^a =\bb\sigma_3$ for all $a=1,\dots, L$. The hyper-parameters are $\theta=0.1$ (in both cases), while $\beta=5$ and $\lambda =0.25$ for the pairwise model (according to \cite{N_of_NN}), while for the dense case $\beta=10$ and $\lambda=3.5$, see Fig. \ref{fig:TAM-phase-diag}: these values of the parameters force the networks to operate in their disentangling regions, thus ensuring a fair comparison between models (as already noted, comparing models at identical values of the hyperparameters does not generally provide a meaningful benchmark, since the effective scales of the thermal level $\beta$ and the anti-imitative strenght $\lambda$ depends on the interaction orders). Neural dynamics is performed in a parallel schedule.}
    \label{fig:dynamics_example}
\end{figure}

As a preliminary check, Fig. \ref{fig:dynamics_example} considers $\alpha=0.4$, $\theta=0.1$ and $L=3$, and shows that the dense architecture ($P=4$ and $D=8$, lower panels) achieves pattern disentanglement from mixture states, whereas the pairwise paradigm ($P=D=2$, upper panels) fails \cite{N_of_NN}. For both architectures, the hyper-parameters $\beta$, $\theta$, and $\lambda$ are chosen so as to place the networks in the disentangling regions of their respective phase diagrams.
For each module, we keep track of the magnetization $m_\mu ^a(t)$ as a function of the update steps. The pairwise setting fails at the task, as the correlation between the neural configurations of the various modules and the corresponding patterns to be retrieved (one per module) is immediately lost. The dense model, on the other hand, achieves correct disentanglement of the three patterns involved in the mixture. Moreover, the corresponding Mattis magnetizations saturate to one within relatively few updates, even in the presence of a linearly extensive number of stored patterns.

\begin{figure}[h]
    \centering
    \includegraphics[width=0.48\linewidth]{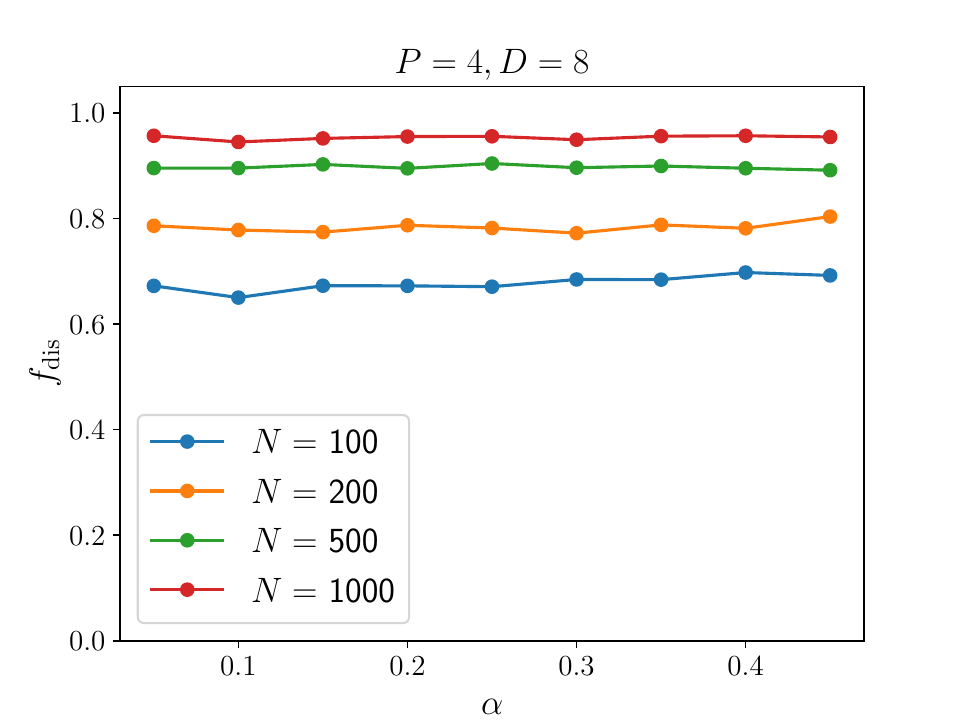}
    \includegraphics[width=0.48\linewidth]{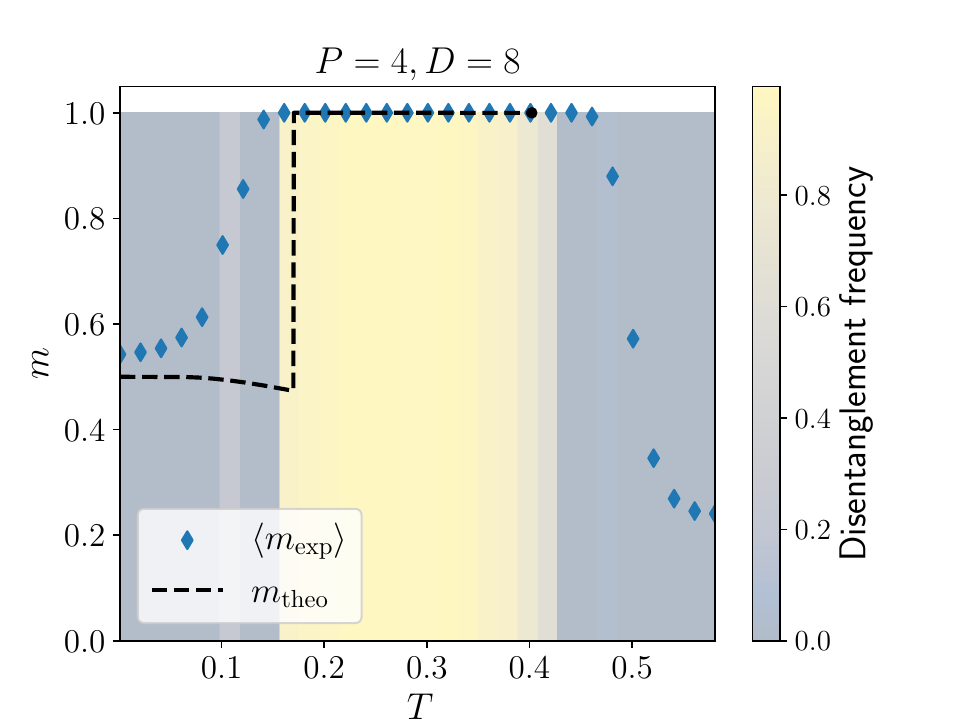}
    \caption{{\bfseries Summary of results of numerical simulations.} (Left) the plot reports the disentangling frequency of the dense model with $P=4$, $D=8$ and $L=3$ starting from a 3-mixture configuration ${\bb\sigma_3} = \operatorname{sign}({\bb\xi^1+\bb\xi^2+\bb\xi^3})$. The results are reported as a function of the ratio $\alpha=K/N$ for various module size $N$. The parameters of the model are chosen so that the model operates in the disentangling phase: $\beta=10$, $\lambda=3.5$, $\theta=0.1$. (Right) The plot reports the module-wise average magnetization (blue points) and the disentangling frequency (color-map) as a function of the temperature (thus moving on a vertical line in Fig. \ref{fig:TAM-phase-diag}, right plot, in order to cross the disentangling region) starting from the 3-mixture configuration ${\bb\sigma_3} $. The former is compared with the stable solutions of the self-consistency equations (Eq. \eqref{eq:self_TAM_dense}) with non-zero magnetization (black dashed line). The model parameters are $\lambda=3.5$, $\theta=0.2$, $N=1000$, $K/N=0.5$. The quantities are computed out of 100 different realizations of the neural dynamics, and evolution is stopped once that the temporal average of module-wise magnetization settles on a constant value.}
    \label{fig:dense_disent}
\end{figure}

As a refined experiment, in Fig. \ref{fig:dense_disent} we better exploit the disentanglement capabilities of the dense model in the linearly extensive setting $K=\alpha N$ as $\alpha$ is made to vary. In the left plot, we investigate the effect of the storage on the pattern extraction performances from the mixtures states for different values of the module's size $N$, with a choice of $(\beta,\lambda,\theta)$ on the boundary of the disentangling region (the yellow phases in Fig. \ref{fig:TAM-phase-diag}). This is assessed by the disentangling frequency $f_{dis}$, defined as the number of successes (in performing disentanglement) over then number of trials. As expected, this is insensitive to the load $\alpha$, its value being constant over a wide range of $\alpha$ values: this is a direct consequence of our choices of {\em densification} of the network and the relative low-load setting. Further, increasing the network size results in $f_{dis}$ approaching one, thus signaling the occurrence, in the thermodynamic limit, of a sharp transition towards an effective disentangling scenario. In the right plot, we instead fix $\lambda$, $\theta$ and $\alpha$ and vary the  temperature (thus moving on a vertical line in the phase diagram in Fig. \ref{fig:TAM-phase-diag}) and compute both the Mattis magnetization related to a stored pattern involved in the starting mixture state (left axes) and the disentangling frequency (right axes). At extremely low temperatures, escaping spurious states is prohibitive thus, even if the neural dynamics takes place, the final magnetization will stay close to the overlap of the three-mixture states ($m\approx 0.5$); at contrary, increasing the temperature yields both an increase of the disentangling frequency and an improve in  the accuracy of the retrieval ($m \approx 1$). A further increase of the thermal noise results in a gradual degradation of the reconstruction capabilities of patterns involved in mixture states. We stress that this whole picture is compatible with the solution of the self-consistency equations \eqref{eq:self_TAM_dense}. 

\section{Applications to Cryptography}\label{sec:apps}
 
We now explore two applications inspired by practical problems in cryptography.

\subsection{A toy decoding problem: pattern reconstruction from Hebbian tensors}
\label{sec:patt_rec}

In the first application, we regard the mixture of patterns as an encrypted message, so that the network's task is to decode it, namely to recover the individual patterns combined therein. Equivalently, this can be framed as an inference problem: the prior information available to the network consists of the $n$-body correlations among the patterns, stored in the Hebbian tensors, together with a collection of $d$ observed mixtures
\begin{equation}
\bb x^\gamma = \operatorname{sign}\Big(\sum_{\mu=1}^K z_\mu^{\gamma} \bb\xi^\mu\Big), \qquad \gamma=1,\dots,d,
\end{equation}
where the coefficients $z^\gamma_\mu \in \{0, 1\}$ are unknown random variables. The goal is to reconstruct the $K$ hidden patterns $\bb\xi^1,\dots,\bb\xi^K$ using only $\{\bb x^\gamma\}_{\gamma=1}^d$ and the Hebbian tensors as input (\emph{vide infra}).

As shown in \cite{ICLR_TAM} for the pairwise architecture ($n=2$), this task can be accomplished through a three-step procedure:
\begin{enumerate}
    \item[\it i)] \emph{Candidate generation.} Each of the $d$ mixtures $\bb x^\gamma$ is used, in turn, as the initial condition for every module of the network, and the neural dynamics is run to convergence. The resulting thermalized configurations, collected across all layers and all mixtures, form a pool of candidate reconstructed patterns.
    \item[\it ii)] \emph{Redundancy removal.} Since different modules, or different runs, may converge to the same pattern, candidates are pairwise compared via their mutual overlap, and near-duplicates are discarded, retaining only one representative per distinct pattern.
     \item[\it iii)] \emph{Acceptance test.} The surviving candidates are filtered through an acceptance criterion, built from the pairwise Hebbian tensor 
$\bb J^{(2)}$, that checks their compatibility with the stored correlations and discards spurious configurations that do not correspond to genuine ground-truth patterns, see \cite{ICLR_TAM,bumi} for more details.
\end{enumerate}
One can further estimate, a priori, the minimal number of mixtures $d_{\min}$ needed to guarantee a satisfactory reconstruction \cite{ICLR_TAM}: this scales as $d_{\min} \sim \frac{K}{L}\log\frac{K}{\epsilon}$,
where $\epsilon$ is the confidence level of the reconstruction, i.e. $\mathrm{Prob}(\text{all } \bb\xi^\mu \text{ reconstructed}) = 1-\epsilon$.

\begin{figure}[h]
    \centering
    \includegraphics[width=0.48\linewidth]{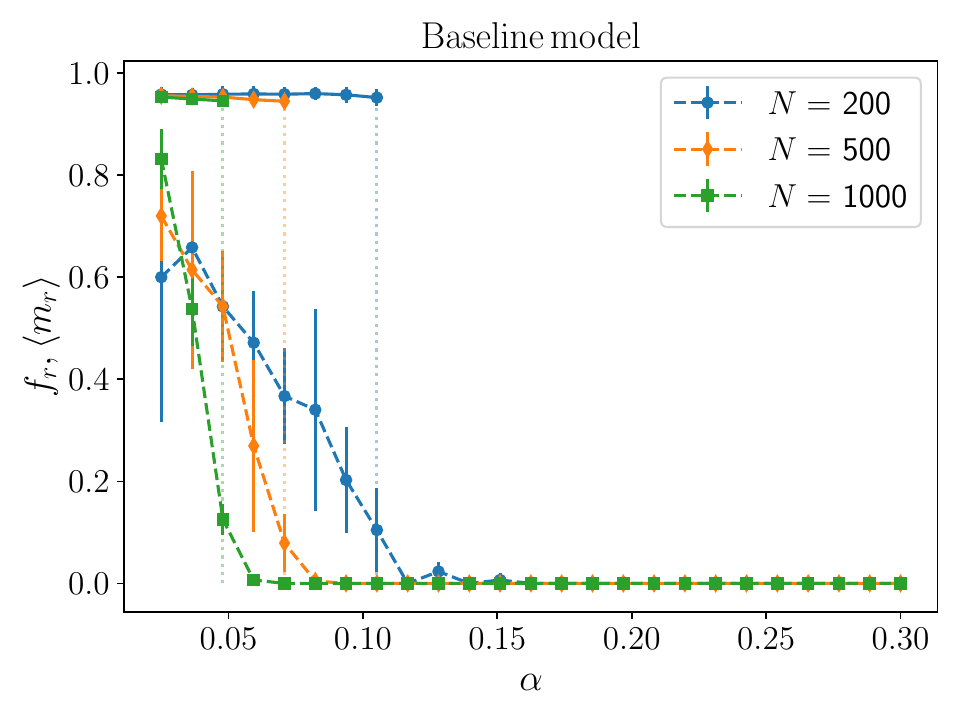}
    \includegraphics[width=0.48\linewidth]{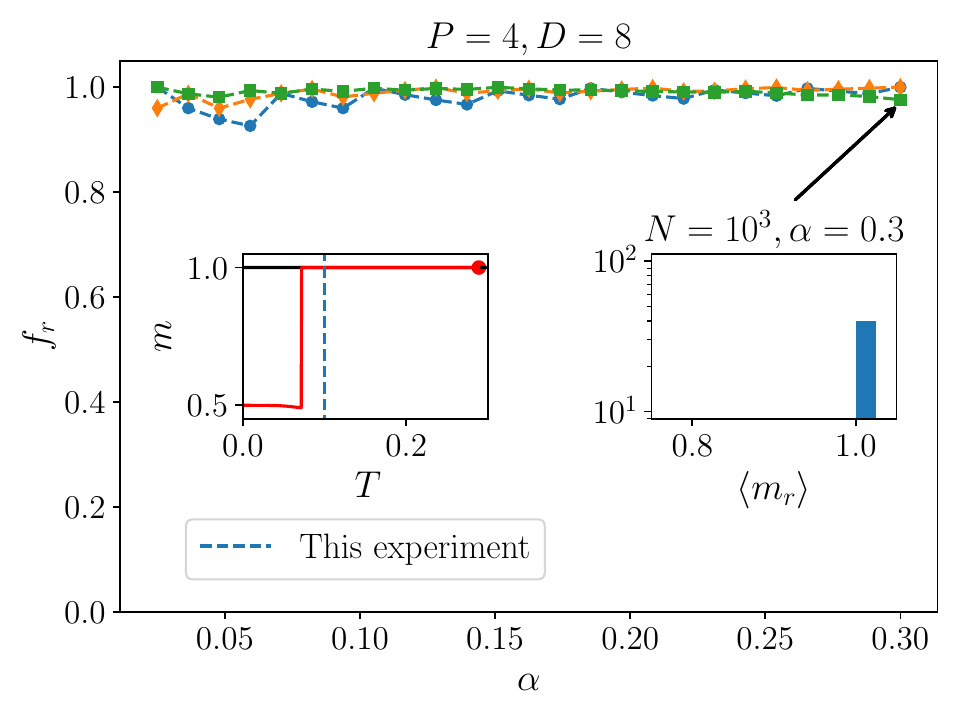}
    \hfill
   \caption{{\bfseries Pairwise {\it vs} dense pattern reconstruction.} Left: pairwise model ($P=D=2$) of \cite{ICLR_TAM}. We report the reconstruction frequency $f_r$ (dashed curves), i.e., the fraction of stored patterns that are correctly reconstructed, and the mean overlap $\langle m_r\rangle$ between reconstructed and ground-truth patterns (solid curves) as functions of the load $\alpha=K/N$, for different network sizes $N$; the overlap is only reported when $f_r>0.05$. Hyper-parameters: $\beta=2$, $\lambda=0.2$, $\theta=0.1$ (taken from \cite{ICLR_TAM}). Right: dense model ($P=4$, $D=8$), reconstruction frequency $f_r$ for $N=200,500,1000$. Hyper-parameters: $\beta=10$, $\lambda=3.5$, $\theta=0.1$. Left inset: stable solution of the self-consistency equations \eqref{eq:self_TAM_dense} as a function of $T=1/\beta$; the vertical dashed line marks the temperature used in the experiment. Right inset: histogram of the overlaps between reconstructed and ground-truth patterns for $N=1000$ and $\alpha=0.3$. In both panels, $L=3$ and the number of observations is $d=2d_{min}$, with $d_{min}=\frac KL\log\frac K\epsilon$ and $\epsilon=0.05$. The dynamics is run for $500$ parallel updates.}
    \label{fig:pattern_reconstruction}
\end{figure}
This algorithm was originally confined to the low-storage regime $K/N \ll 1$, where the pairwise Hebbian tensor $\bb J^{(2)}$ still carries enough information to disentangle the mixtures reliably. In the linearly extensive regime $K = \Theta (N)$, this information is no longer sufficient: pairwise correlations become too noisy to isolate individual patterns. Dense extensions of the network, operating in their own low-load regime (as explained in the first part of this manuscript), overcome this limitation because $n$-point correlations -- Hebbian-stored in the higher-order auto-associative tensors -- retain much richer information about the hidden patterns than pairwise correlations alone.

Concretely, for $P=4, D=8$, the prior knowledge available to the dense network is encoded in the tensors $\bb J^{(4)}$ and $\bb J^{(8)}$; the pairwise tensor $\bb J^{(2)}$ is still needed only to implement the acceptance criterion of step \emph{iii)} \cite{ICLR_TAM,bumi}.

Figure~\ref{fig:pattern_reconstruction} compares the reconstruction performance of the pairwise baseline and its dense extension under matched conditions -- same scaling regime $K=\alpha N$ and same input mixtures $\bb x^\gamma$ -- with hyper-parameters chosen, in both cases, to place the network in its disentangling phase. Performance is quantified via two indicators: the reconstruction frequency $f_r$, i.e. the fraction of correctly recovered patterns, and the mean overlap $\langle m_r \rangle$ between the accepted, reconstructed candidates and the ground-truth patterns.
The pairwise model's reconstruction frequency degrades rapidly as the load $\alpha$ increases, an effect that becomes more severe as $N$ grows -- indicating that this baseline is reliable essentially only at $\alpha \to 0$. The dense model, by contrast, retains robust reconstruction frequency and retrieval quality across the whole range of $\alpha$ and $N$ tested, confirming that higher-order correlations are what enables reconstruction in the extensive storage regime.

\subsection{A cryptographic Proof-of-Concept}\label{sec:crypto}

\begin{figure}[t]
    \centering
    \includegraphics[width=15cm]{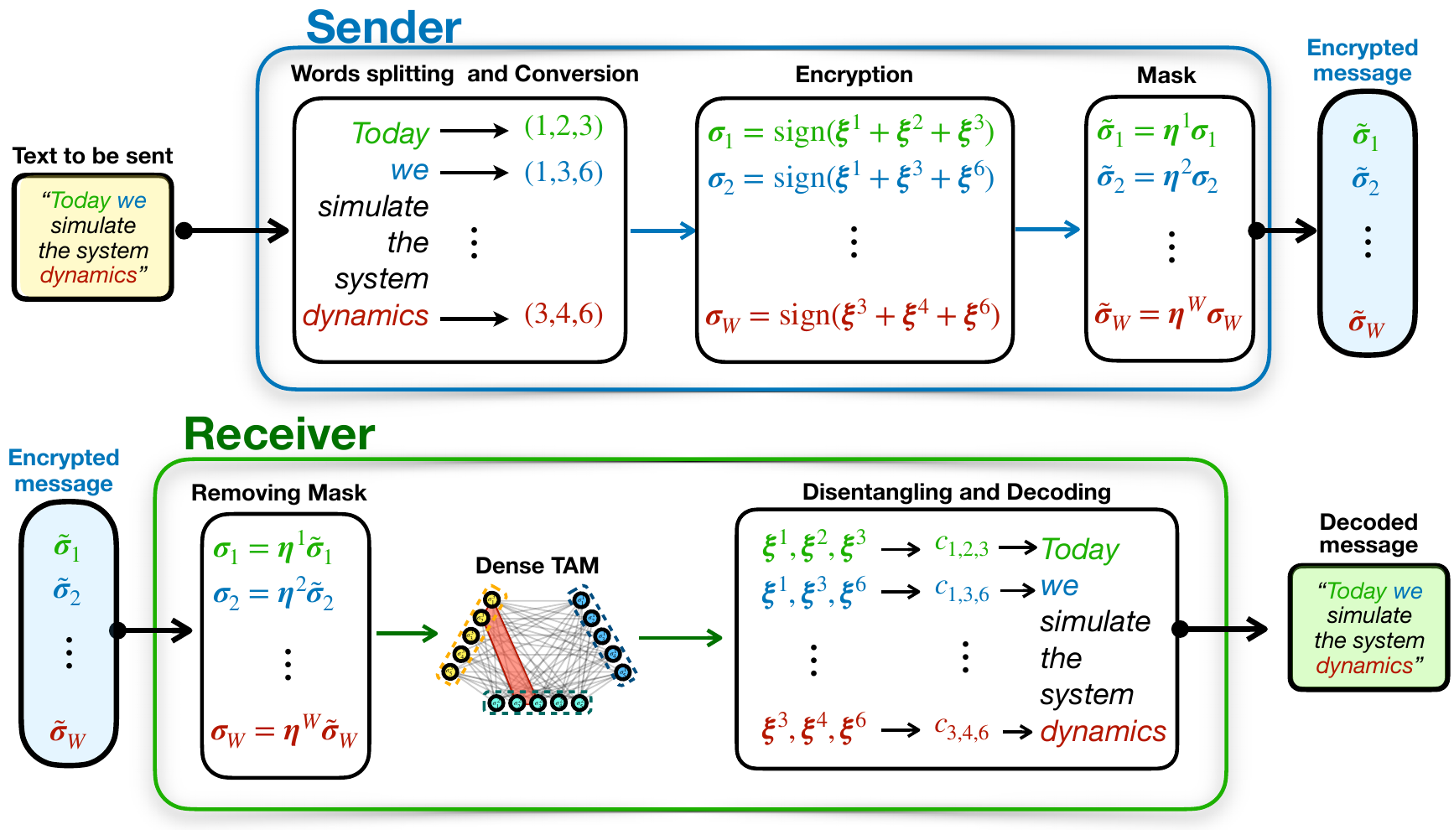}
    \caption{{\bfseries Schematic representation of the modular associative memory based encryption/decryption protocol} (online stage). After the initialization stage, sender and receiver share the hidden-pattern dictionary. The sender splits the plaintext into lexical tokens and maps each of them to a triplet of hidden-pattern indices, e.g., $(1,2,3)$, $(1,3,6)$ or $(3,4,6)$. Each triplet is encoded as a spurious mixture state, $\bm\sigma_w=\operatorname{sign}(\bm\xi^{\mu_w}+\bm\xi^{\nu_w}+\bm\xi^{\ell_w})$ (the encoding noise $T_{enc}\tilde{\bm u}$ of Eq.~\eqref{eq:starting_with_noise} is omitted for clarity), and masked through a fresh pseudorandom vector $\bm\eta^w$, see Eq.~\eqref{eq:mask}, yielding the transmitted ciphertext $\tilde{\bm\sigma}_w=\bm\eta^w\odot\bm\sigma_w$. The encrypted message is thus a sequence of masked spurious states. The receiver regenerates and removes the masks, recovering the mixture states $\bm\sigma_w$, and feeds each of them to the dense TAM (i.e. three directional associative memory), whose disentangling dynamics retrieves the three constituent hidden patterns. These are combined into the Boolean code $\bm{c}_{\mu\nu\ell}$ of Eq.~\eqref{eq:code}, which is matched against the shared codebook, thereby reconstructing the plaintext.}
    \label{fig:toymodel_crypto}
\end{figure}

We now turn disentanglement into a communication protocol, sketched in Fig.~\ref{fig:toymodel_crypto}. The idea is simple: each lexical token\footnote{By \emph{token} we generically mean a word, a punctuation mark or a space.} is associated with a triplet of hidden patterns and is transmitted as the corresponding (noisy) three-pattern mixture, further concealed by a pseudorandom mask. The legitimate receiver, who owns the synaptic tensors of a dense TAM  (i.e. three directional associative memory), removes the mask and lets the network disentangle the mixture: the three modules converge to the three constituents, which identify the token. The choice $L=3$ is thus natural, as the number of modules matches the number of patterns per token; furthermore, three-pattern mixtures are the most stable spurious states of Hopfield-like networks \cite{CKS}, so that they are well-defined, reproducible objects to be transmitted. Two features make this scheme appealing. First, the ciphertext is a mixture, that is, exactly the kind of object the network is designed to undo. Second, the decoding relies on attractor dynamics, which is intrinsically error-correcting: a corrupted mixture still lies in the basin of attraction of the correct disentangled state, so that robustness against channel noise comes for free.
\newline
The protocol consists of an \emph{initialization} stage, performed only once, and an \emph{online} stage, repeated for each transmitted token.

\paragraph{Initialization.} The sender draws $K$ Rademacher patterns $\bm{\xi}^\mu\in\{-1,+1\}^N$, $\mu=1,\dots,K$. These carry no semantic content: they play the role of microscopic secret keys. Each token of a vocabulary of size at most $\binom{K}{3}$ is associated with a triplet of distinct indices $(\mu,\nu,\ell)$, with $\mu,\nu,\ell\in\{1,\dots,K\}$; the resulting dictionary $\texttt{token}\mapsto(\mu,\nu,\ell)$ can even be public. Next, each triplet is associated with the Boolean code
\begin{equation}
\texttt{token} \longmapsto (\mu,\nu,\ell) \longmapsto \bm{c}_{\mu\nu\ell}=\Theta(\bm\xi^\mu\odot\bm\xi^\nu\odot\bm\xi^\ell)\in \{0,1\}^{N},
\label{eq:code}
\end{equation}
where $\Theta$ is the Heaviside function and $\odot$ the entry-wise product. The reason for this choice is that the disentangling dynamics returns the three constituents on the three modules in an arbitrary order (all the $3!$ assignments are equally good disentangled states): the receiver thus needs an identifier that is invariant under permutations of the triplet, and the entry-wise product is the simplest one. Moreover, codes of distinct triplets are nearly orthogonal (their mutual overlap, as $\pm1$ vectors, is $\mathcal O(N^{-1/2})$), so that matching a noisy code against the codebook is robust. Unlike the dictionary, the codebook $\texttt{token}\mapsto\bm c_{\mu\nu\ell}$ depends on the secret patterns and is shared with the receiver only. Finally, the receiver is provided with the Hebbian tensors $\bb J^{(2)}$, $\bb J^{(P)}$ and $\bb J^{(D)}$, and with the seed of a pseudorandom generator used for masking (see below). Codebook, tensors and seed constitute the pre-shared secret key; the patterns themselves are never transmitted. Notice that, as shown in Sec.~\ref{sec:patt_rec}, the tensors do allow for reconstructing the patterns: the secrecy of the protocol relies on keeping the whole pre-shared material private, not on hiding the patterns from the legitimate receiver.

\begin{figure}[t]
    \centering
\includegraphics[width=15cm]{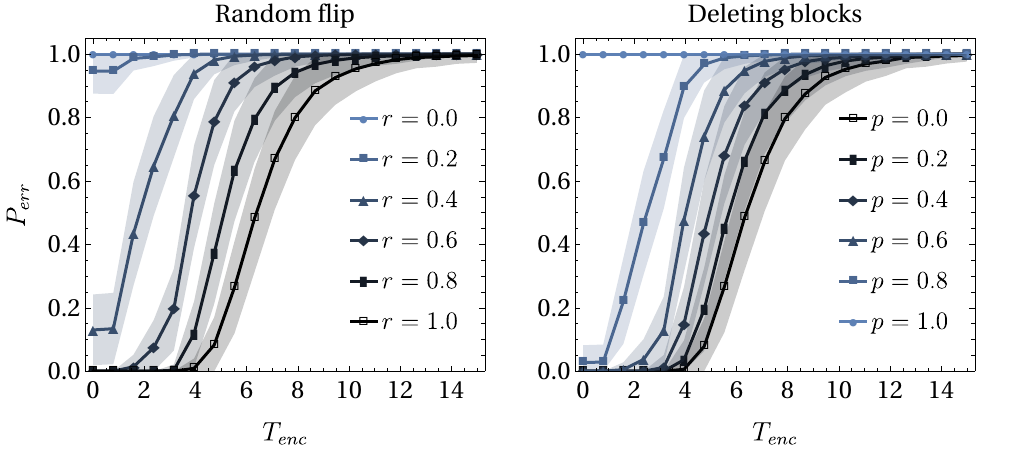}
    \caption{{\bfseries Decoding performance of the modular associative memory based protocol under channel corruption.} Word-reconstruction error probability $P_{err}$ as a function of the encoding temperature $T_{enc}$, for random flips (left, different values of $r$, where each entry is flipped with probability $(1-r)/2$, so that $r=1$ is the noiseless channel) and block erasures (right, different values of the erased fraction $p$). On a clean channel, decoding is essentially error-free up to $T_{enc}\approx 4$; corruption shifts this threshold toward lower encoding temperatures. Decoding fails at all encoding temperatures for flip fractions of $40\%$ or more ($r\le 0.2$), whereas under block erasures only the complete erasure ($p=1$) prevents decoding.  Each point is obtained by simulating the transmission of $10^4$ tokens with $N=500$, $K=90$, $L=3$, $P=4$, $D=8$, $\beta=10$, $\lambda=3.5$, and $\theta=0.1$.}
    \label{fig:crypto_performance}
\end{figure}

\paragraph{Encryption.} To transmit the token $(\mu,\nu,\ell)$, the sender builds the noisy spurious state
\begin{equation}
    \bm{\sigma}_{\mu\nu\ell} = \operatorname{sign}\!\left(\bm{\xi}^\mu+\bm{\xi}^\nu+\bm{\xi}^\ell + T_{enc}\tilde{\bm{u}}\right),
    \label{eq:starting_with_noise}
\end{equation}
where $\tilde{\bm{u}}$ is a zero-mean random vector with compact support, drawn afresh at each transmission (in our experiments, $\tilde{\bm{u}}\sim\mathcal U([-1,1]^N)$), and the \emph{encoding temperature} $T_{enc}\ge 0$ tunes its amplitude. For $T_{enc}=0$, a given token always produces the same vector; the noise, instead, makes different occurrences of the same token different, and weakens the correlation between the transmitted state and each of its constituents. Quantitatively, the site-wise correlation $\mathbb E[\sigma_{\mu\nu\ell,i}\,\xi^\mu_i]$ equals $1/2$ for $T_{enc}\le 1$ and decays as $1/T_{enc}$ for $T_{enc}\ge 3$ (see App.~\ref{app:corr_input}). Moderate values of $T_{enc}$ also help the decoder, as they destabilize the input mixture and favor the escape toward the disentangled configuration, similarly to the beneficial role of thermal noise discussed in Sec.~\ref{sec:numerics}. Too large values, however, erase the correlation with the stored patterns and prevent disentanglement. The encoding temperature thus rules a trade-off between obfuscation and decodability, which is explored in Fig.~\ref{fig:crypto_performance}.
\newline
The state $\bm{\sigma}_{\mu\nu\ell}$ still carries exploitable statistical regularities: it is strongly correlated with three patterns and, over a long message, it inherits the frequency statistics of the underlying language. To conceal them, for each transmitted token $w$ the sender draws a fresh pseudorandom Rademacher vector $\bm\tau^w\in\{-1,+1\}^N$ from the shared seed, and processes it through a single step of Hopfield dynamics driven by $\bb J^{(2)}$:
\begin{equation}
{\eta}^w_i = \operatorname{sign}\Big(\sum_{\substack{j=1 \\ j\neq i}}^{N}J^{(2)}_{ij}\,\tau^w_j\Big), \qquad i=1,\dots,N.
\label{eq:mask}
\end{equation}
The transmitted ciphertext is then
\begin{equation}
    \tilde{\bm{\sigma}}_{\mu\nu\ell} = \bm{\eta}^w \odot \bm{\sigma}_{\mu\nu\ell},
    \label{eq:cipher}
\end{equation}
which, in Boolean representation, corresponds to a bitwise XOR: the sequence of masks $\bm\eta^w$ plays the role of a keystream. Using $\bm\eta^w$ rather than $\bm\tau^w$ makes the mask depend on two independent secrets, the seed and the tensor $\bb J^{(2)}$: an attacker who learned only one of them could not compute the mask. The effectiveness of this layer against frequency analysis and against learning-based inference attacks is assessed in App.~\ref{app:masking}: after masking, the rank--frequency statistics of the ciphertexts is flat, and a classifier trained on (ciphertext, plaintext) pairs performs at chance level.

\paragraph{Decryption.} The receiver regenerates $\bm\eta^w$ from the shared seed and, since $(\eta^w_i)^2=1$, recovers $\bm\sigma_{\mu\nu\ell}=\bm\eta^w\odot\tilde{\bm\sigma}_{\mu\nu\ell}$. This state is used both as initial condition and as external field for all the modules, $\bm\sigma^a(t=0)=\bm h^a=\bm\sigma_{\mu\nu\ell}$ for $a=1,2,3$, as in Sec.~\ref{sec:numerics}, and the dynamics \eqref{eq:update_MC_iclr} is run with hyper-parameters in the disentangling phase. At convergence, the three modules are aligned with the three constituents $\bm\xi^\mu,\bm\xi^\nu,\bm\xi^\ell$ (in some order); the receiver computes $\hat{\bm c}=\Theta(\bm\sigma^1\odot\bm\sigma^2\odot\bm\sigma^3)$ and outputs the token whose code is closest to $\hat{\bm c}$ in Hamming distance. A message made of $W$ tokens is thus transmitted as the $W\times N$ matrix obtained by stacking the corresponding ciphertexts, and is decoded token by token. Notice that the receiver never needs to handle the patterns explicitly: the dynamics only uses the tensors, and the token is identified through its code.

\paragraph{Channel corruption.} Between sender and receiver, the ciphertext may be corrupted, either by physical noise or by a third party tampering with the channel. We model the received vector as
\begin{equation}
\hat{\bm\sigma}_{\mu\nu\ell}=\bm\chi\odot\tilde{\bm\sigma}_{\mu\nu\ell},
\label{eq:corruption}
\end{equation}
and consider two corruption models:
\begin{itemize}
    \item \emph{Random flips}: each entry is flipped independently with probability $(1-r)/2$, that is,
\begin{equation}
    \mathcal{P}(\chi_i)= \frac{1+r}{2}\delta(\chi_i-1)+ \frac{1-r}{2}\delta(\chi_i+1), \qquad r\in [0,1],~~i=1,\dots,N;
    \label{eq:random_flip}
\end{equation}
    $r=1$ corresponds to a noiseless channel, while $r=0$ to a completely random output.
    \item \emph{Block erasures}: a contiguous block containing a fraction $p$ of the entries is cleared, i.e., set to $0$. For a starting index $k\in\{1,\ldots,N-1\}$ and a block fraction $p\in[0,1]$,
\begin{equation}
    \chi_i =
    \begin{cases}
    0 & \text{if }\;\; k \leq i \leq \min(k + pN, N) \\
    1 & \text{otherwise}
    \end{cases}.
    \label{eq:gaussian_noise}
\end{equation}
\end{itemize}
Both corruptions act entry-wise and, therefore, commute with the mask: after unmasking, the receiver holds $\bm\eta^w\odot\hat{\bm\sigma}_{\mu\nu\ell}=\bm\chi\odot\bm\sigma_{\mu\nu\ell}$. The robustness of the protocol thus coincides with the robustness of the disentangling dynamics against a corrupted input. Moreover, since all sites are statistically equivalent for the network (the patterns are i.i.d.\ across sites and the Hebbian tensors are permutation invariant), a contiguous block of erasures is exactly as harmful as the same number of scattered erasures: unlike conventional channel codes, the protocol is insensitive to the burst structure of the corruption.

\paragraph{Performance.} The decoding performance is summarized in Fig.~\ref{fig:crypto_performance}, where we report the word-reconstruction error probability $P_{err}$ as a function of the encoding temperature. On a clean channel ($r=1$ or $p=0$), decoding is essentially error-free up to $T_{enc}\approx 4$, beyond which the encoding noise destroys the correlation with the constituent patterns and $P_{err}$ increases toward one. Channel corruption shifts this threshold toward lower encoding temperatures, and the protocol turns out to be much more tolerant to erasures than to flips, as an erased entry carries no information, whereas a flipped one carries wrong information. With $60\%$ of erased entries ($p=0.6$), decoding is still essentially error-free up to $T_{enc}\approx 2.5$; even with $80\%$ of erased entries ($p=0.8$), the error probability is of a few percent at small $T_{enc}$, and only a complete erasure ($p=1$) prevents decoding. Under random flips, decoding is essentially error-free up to $T_{enc}\approx 2$ with $20\%$ of flipped entries ($r=0.6$); with $30\%$ ($r=0.4$) the error probability is about $13\%$ already at small $T_{enc}$, and for $40\%$ or more ($r\le 0.2$) decoding fails at all encoding temperatures. In practice, $T_{enc}$ should thus be set as large as allowed by the expected channel quality, so as to maximize obfuscation while keeping $P_{err}$ negligible. Residual decoding errors can be further suppressed at the message level by repeating the whole transmission a finite number of times: assuming independent token-wise errors, this allows trading communication cost for reliability in a controlled way, as detailed in App.~\ref{app:repetitions}.

\paragraph{Comparison with standard pipelines.} To put these results in context, we compare the protocol with two conventional pipelines for secure transmission over noisy channels: AES-GCM encryption \cite{AES1,AES2} followed by LDPC channel coding \cite{LDPC1,LDPC2} (AES+LDPC), and a post-quantum pipeline based on the ML-KEM (Kyber) key-encapsulation mechanism \cite{Kyber1,Kyber2}, again protected by LDPC coding (Kyber+LDPC). The three schemes are tested on the same message-transmission task and under the same corruption models, using $P_{\mathrm{err}}$ as performance metric. We stress that this is not a like-for-like comparison of security. AES-GCM and ML-KEM are designed to guarantee integrity: any alteration of the ciphertext that survives channel decoding causes, by construction, the rejection of the whole block (failed authentication or wrong decapsulated key). Our scheme offers no integrity guarantee and aims instead at recovering the semantic content of the message. The conventional pipelines should therefore be regarded as reference baselines for message recovery under signal corruption, not as competitors in terms of security.

\begin{figure}[!t]
    \centering
    \includegraphics[width=\textwidth]{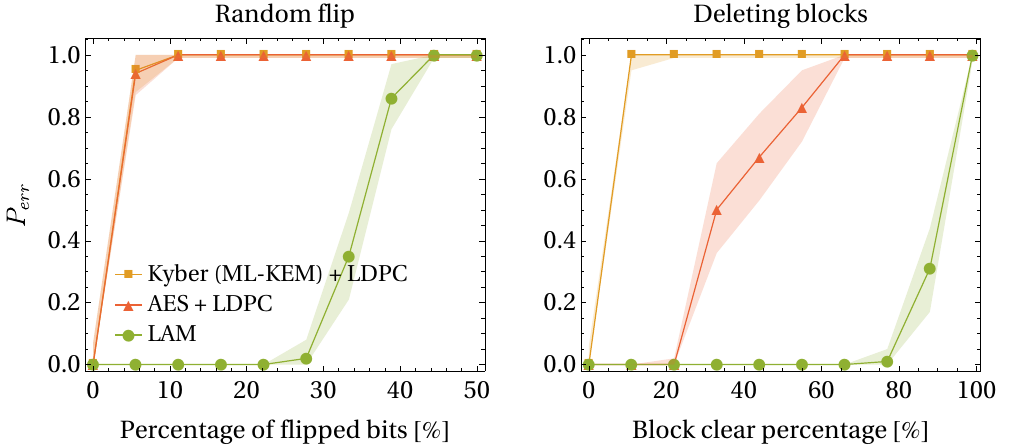}
     \caption{Comparison between the proposed associative-memory-based communication scheme and conventional secure-transmission baselines (AES+LDPC and Kyber (ML-KEM)+LDPC) on noisy channels. Left panel: resilience to random spin flips, measured through the word-reconstruction error probability ($P_{err}$) as a function of the percentage of flipped bits. Right panel: resilience to structured cancellations (signal holes), quantified by the $P_{err}$ as a function of the erased/cleared block fraction. In both settings, the associative-memory-based protocol retains correct decoding over a substantially broader noise range, showing markedly more graceful degradation under severe channel corruption. The associative memory works with the same hyperparameters used in Fig.~\ref{fig:crypto_performance}, with $T_{enc}=0$.}
    \label{fig:crypto_vs_literature}
\end{figure}
We quantify decoding performance using the word-reconstruction error probability $P_{\mathrm{err}}$, defined as the empirical fraction of independent input realizations (i.e., independently sampled transmitted words and channel noise instances) for which the receiver's output does not coincide with the transmitted lexical item. As shown in Fig.~\ref{fig:crypto_vs_literature}, the modular associative memory  protocol displays a substantially wider robustness region under both kinds of corruption. Under random flips, its $P_{\mathrm{err}}$ remains essentially zero up to approximately $25\%$--$30\%$ of flipped bits and then increases gradually, whereas both conventional pipelines display the typical cliff effect, failing almost completely already for a small percentage of corrupted bits. A similar picture emerges under block erasures: the modular associative memory  protocol remains essentially error-free up to about $80\%$ of cleared entries, whereas the Kyber-based pipeline collapses almost immediately and the AES+LDPC one starts degrading at intermediate erasure levels and fails completely well before the associative-memory scheme.
\newline
This difference stems from the different decoding principles. In the conventional pipelines, decoding is bitwise: once the channel errors exceed the correction capability of the LDPC code, residual errors reach the cryptographic layer, which by design rejects the whole block; furthermore, contiguous erasures are particularly harmful for channel codes. In the modular associative memory  protocol, decoding is collective: the received state does not need to coincide with the transmitted one, but only to lie in the basin of attraction of the correct disentangled configuration, and, as noted above, the position of the corrupted entries is irrelevant. We also note that the present modular associative memory protocol spends $N=500$ binary channel symbols per token, while a token carries at most $\log_2\binom{K}{3}\approx 17$ bits for $K=90$, i.e., its code rate is $\approx 0.034$. Part of its robustness thus stems from this redundancy: we point out that the Shannon limit of the binary symmetric channel at this rate corresponds to a flip fraction $\approx 39\%$, close to the breakdown of our protocol.
\newline
The robustness reported in Fig.~\ref{fig:crypto_vs_literature} should therefore be interpreted as an application-level advantage in noisy or adversarially disturbed channels, where graceful semantic recovery is preferable to all-or-nothing bitwise integrity, and not as a claim of cryptographic superiority over standard, formally verified protocols. In the present proof of concept, confidentiality relies on the secret patterns, on the pre-shared codebook and tensors, and on the pseudo-random masking layer, which numerically suppresses frequency-based and data-driven inference attacks (App.~\ref{app:masking}); a full information-theoretic or computational security analysis is beyond the scope of this work. Within these limits, the protocol provides a compelling option whenever the physical channel is extremely unreliable and graceful degradation takes precedence over exact symbol-by-symbol recovery.

\section{Conclusion}
Associative memories have traditionally been viewed as devices that retrieve stored information from partial or noisy cues. In this work we have shown that, when several dense associative memories interact through suitable hetero-associative couplings, their collective behavior gives rise to a qualitatively different functionality -- more is different \cite{Anderson72}: rather than simply recognizing a stored pattern, the network can split a composite signal into its elementary constituents and retrieve each of them on a different module. Spurious mixtures, historically regarded as the Achilles' heel of Hopfield networks, thus become the input of a well-defined computational task.

To investigate this phenomenon, we introduced a dense modular associative memory in which both the auto-associative couplings within each module and the hetero-associative (anti-imitative) couplings among modules are promoted to higher-order Hebbian interactions, of order $P$ and $D$, respectively. Choosing $2<P\le D$ suppresses both sources of slow noise, the one induced by the stored patterns and the one induced by the other modules, so that the network operates in a low-load regime even when the number of stored patterns scales linearly with the module size, $K=\alpha N$. This allowed us to carry out the statistical-mechanical analysis in terms of Mattis magnetizations only, via Guerra's interpolation under the replica-symmetric ansatz, and to draw phase diagrams where a disentangling region, in which spurious mixtures are unstable while disentangled configurations are stable, is clearly identified. Unlike the pairwise case, the boundary of this region does not depend on the load and, for $P>2$, the disentangling-to-ergodic transition is discontinuous. Monte Carlo simulations confirm this picture and show that the dense network disentangles three-pattern mixtures at loads where its pairwise counterpart fails.

We then illustrated how disentanglement can be turned into a decoding primitive. First, we showed that the full set of hidden patterns can be reconstructed explicitly from the Hebbian tensors and from a stream of unlabeled mixtures; the dense algorithm succeeds at extensive load, where the pairwise one only works at vanishing load. Second, we built a proof-of-concept communication protocol in which each lexical token is encoded as a noisy three-pattern mixture, concealed by a pseudorandom mask, and decoded by the disentangling dynamics of a dense  three directional associative memory. Since decoding is carried out by attractor dynamics, the received state only needs to fall in the basin of attraction of the correct disentangled configuration: the protocol therefore degrades gracefully under random flips and block erasures, and it is insensitive to the burst structure of the corruption. This robustness should be read as an application-level advantage over very unreliable channels, not as a claim of cryptographic security, whose formal analysis is left for future work.

Our focus on $L=3$ is motivated both theoretically and phenomenologically. On the theoretical side, three-pattern mixtures are the most stable symmetric spurious states of Hopfield-like networks \cite{CKS}. On the phenomenological side, several natural signals are superpositions of three components: a color image is the superposition of three primary color channels, and a major triad is the superposition of three notes (e.g., C, E and G for the C major chord). Other small values of $L$ are equally meaningful: $L=2$ corresponds to the cocktail party problem, i.e., the core of blind source separation; $L=4$ adds an opacity channel to the color picture; $L=5$ could mimic multisensory integration, where modules would separate the contributions of different senses fused into a common representation, a question of interest in cognitive science.

Several directions remain open. On the theoretical side, the saturated regime $K\propto N^{P-1}$, where replica symmetry breaking is expected to play a role, as well as heterogeneous architectures with module-dependent interaction orders $P(a)$ and $D(a,b)$, deserve a dedicated analysis. On the computational side, the dynamics can be implemented through the overlaps $m^a_\mu$, at a cost $\mathcal O(KN)$ per update, without ever storing the $\mathcal O(N^P)$ and $\mathcal O(N^D)$ tensor entries; this, however, requires the patterns explicitly, and it would be interesting to understand whether compact or learned (rather than Hebbian) representations of the couplings can reconcile efficiency with the secrecy requirements of the protocol. On the applicative side, a formal security analysis and a comparison with channel codes at matched rate are natural next steps. More broadly, we believe that the step from pattern recognition to pattern disentanglement opens a promising research direction at the interface of statistical mechanics, neural computation and information theory: the ability of neural networks to disentangle structured information may prove as important as their ability to store it, and this whole class of phenomena turns out to be accessible to the statistical mechanics of neural networks.

\section*{Acknowledgments}

A.B. and E.A. acknowledge financial support from Sapienza University of Rome, project number RM12519999AB8CA9 (“Neural Networks and Learning Machines: asymptotic behaviors on structured datasets”). E.A. and A.F. acknowledge financial support from PNRR MUR project PE0000013-FAIR and from Sapienza University of Rome (RM124190CB1269EB). The Authors are deeply grateful to Dmitry Krotov for the valuable discussions.

\bibliographystyle{unsrt}  
\bibliography{biblio}

\newpage

\appendix

\section{Guerra interpolation, RS ansatz and self-consistency equations}
\label{app:guerra_interpol}
In this Appendix we collect the detailed derivations of the statistical-mechanical solution of the modular associative memory  under consideration. 
In the retrieval regime, we require all the $L$ modules to retrieve (exhaustively) $L$ patterns entering the input mixtures \eqref{eq:init}. Without loss of generality, we may label these as the first $L$ patterns among the $M$ involved in those combinations. 
In this scheme, we can separate the contributions associated with the retrieved (signal) patterns, $1\leq \mu\leq L$, from those associated with the remaining (non-condensed) patterns, $\mu>L$, in the partition function defined in Eq. \eqref{partition-function_TAM_easy}. Then
\begin{equation}
\label{eq:part_splitted_patt1}
\begin{array}{lll}
       \mathcal{Z}_{N,K,L}^{(P,D)}(\beta\,|\,\bm{h},\bm{\xi},\lambda,\theta)&= \SOMMA{\bm\sigma\in \Omega}{} \exp\Big( \dfrac{\beta N}{2} \sum\limits_{a=1}^L\sum\limits_{\mu=1}^{L} (m_{\mu}^a)^P-\beta N \lambda\sum\limits_{\mu=1}^{L} \sum\limits_{a=1}^L\sum_{b>a}^{L}(m_{\mu}^a)^{D/2}(m_{\mu}^b)^{D/2}\\&+\beta \theta \sum\limits_{a=1}^{L}\sum\limits_{i=1}^{N}h_i^a \sigma_i^a
     +\dfrac{\beta}{2 N^{P-1}} \sum\limits_{a=1}^L\sum\limits_{\mu>L}^{K} \sum\limits_{i_1,\cdots, i_P}^{N}\xi_{i_1}^\mu \cdots \xi_{i_P}^\mu \sigma^a_{i_1} \cdots \sigma_{i_P}^a\\&-\dfrac{\beta\lambda}{N^{D-1}} \sum\limits_{\mu>L}^{K} \sum\limits_{a=1}^L \sum\limits_{b>a}^{L} \SOMMA{i_1,\cdots , i_D}{N} \xi_{i_1}^\mu \cdots \xi_{i_D}^\mu \sigma^a_{i_{1}}\cdots\sigma^a_{i_{D/2}}\sigma^b_{i_{D/2+1}}\cdots\sigma^b_{i_{D}}\Big).
\end{array}
\end{equation}
Now, since equilibrium configurations of the model are uncorrelated with non-retrieved patterns, the associated magnetizations are asymptotically Gaussian-distributed with zero mean and variance $1/N$. Let us denote for simplicity $S_\mu ^{(n)}= m_\mu ^n$ with even $n$. For sufficiently large $N$, the probability density of $S_{\mu}^{(n)}$ is well-approximated by the formula
$$
f_n(S) = C_n \sqrt N  S ^{\frac{1}{n}-1}\exp \Big(-\frac N2  S^{2/n}\Big).
$$
From this asymptotic density one readily infers $\mathbb E S_\mu ^{(n)}=N^{-\frac n2}(n-1)!!\doteq \tilde \mu $ and $\text{Var} (S_\mu ^{(n)}) = N^{-n}[(2n-1)!!-((n-1)!!)^2]\doteq \tilde \sigma^2$, see also \cite{agliari2022nonlinear}. Now, the contributions of non-condensed patterns to the energy function involve terms as
\begin{equation}
    \label{eq:largeN}
    \frac{1}{N^{n-1}}\sum_{\mu >L}^K \sum_{i_1,\dots,i_n}^N \xi^\mu_{i_1}\dots \xi^\mu_{i_n}\sigma_{i_1}\dots \sigma_{i_n} = NK \frac1K\sum_{\mu >L } S_{\mu}^{(n)}.
\end{equation}
Since the random variables $S_{\mu}^{(n)}$ are independent for different $\mu$, we can apply the CLT to estimate its order in the regime $K=\alpha N$. In particular, for sufficiently large $N$
$$
\frac1K\sum_{\mu >L } S_{\mu}^{(n)}\sim \mathcal N \big(\tilde \mu, \frac{\tilde\sigma^2}{K}\big).
$$
To understand the typical order of non-condensed patterns contribution, we then consider
$$
NK \tilde \mu = \frac{\alpha N^2}{N^{\frac n2}}(n-1)!! = c_n N^{2-\frac n2},
$$
while
$$
(NK )^2\frac{\tilde\sigma^2}{K}= \frac{\alpha N^3}{N^n}[(2n-1)!!-((n-1)!!)^2] = d_n N^{3-n},
$$
with $c_n,d_n$ being constants independent of $N$. For $n\ge 4$ the variance of contributions in Eq. \eqref{eq:largeN} goes to zero with system size at least as $1/N$, while the expectation is at most finite ($\mathcal O(1)$ only for the $n=4$ case). Then, since $P,D\ge 4$, the net contribution the non-retrieved patterns to the energy function is $\mathcal O(1)$ and, by this, they do not affect the thermodynamic limit of the intensive quenched pressure (i.e. the normalized logarithm of the partition function, {\it vide infra}). As a result, we can safely drop those terms from the partition function, and directly write
\begin{equation}
\label{eq:part_gauss_start}
\begin{array}{lll}
       \mathcal{Z}_{N,K,L}^{(P,D)}(\beta\,|\,\bm{h},\bm{\xi},\lambda,\theta)&\approx \SOMMA{\bm\sigma\in \Omega}{} \exp\Big( \dfrac{\beta N}{2} \sum\limits_{\mu,a=1}^L (m_{\mu}^a)^P-\beta N \lambda\sum\limits_{\mu,a<b=1}^{L} (m_{\mu}^a m_{\mu}^b)^{D/2}+\beta  \theta\sum\limits_{a=1}^{L}\sum\limits_{i=1}^{N}h_i^a \sigma_i^a\Big).
\end{array}
\end{equation}

We proceed to the computation of the quenched free energy given in Eq. \eqref{eq:Free-Definition} via Guerra's interpolation method. Following \cite{guerra}, we introduce a parameter $t \in [0,1]$ and define an interpolating partition function $\mathcal{Z}_{N,K,L}^{(P,D)}(t)$ with associated statistical pressure
\begin{equation}
\label{eq:interpol_pressure}
    \mathcal{A}_{N,K,L}^{(P,D)}(t) = \dfrac{1}{N}\mathbb{E}_{\bm\xi}\log \mathcal{Z}_{N,K,L}^{(P,D)}(t),
\end{equation}
with 
\begin{equation}
\label{eq:interpolating_Z}
\begin{array}{lll}
       \mathcal{Z}_{N,K,L}^{(P,D)}(t) &= \SOMMA{\bm\sigma\in \Omega}{} \exp\Big(t\dfrac{\beta N}{2} \sum\limits_{a=1}^L\sum\limits_{\mu=1}^{L} (m_{\mu}^a(\bm\sigma))^P +N(1-t)\SOMMA{\mu=1}{L}\SOMMA{a=1}{L}\psi_\mu^a m_{\mu}^a(\bm\sigma)+\beta \theta\sum\limits_{a=1}^{L}\sum\limits_{i=1}^{N}h_i^a \sigma_i^a
\\
     &-t \dfrac{\beta N \lambda}{2}\sum\limits_{\mu=1}^{L}  \sum\limits_{{a\neq b}=1}^L(m_{\mu}^a(\bm\sigma) m_{\mu}^b(\bm\sigma))^{D/2} -(1-t)N\sum\limits_{\mu=1}^{L} \sum\limits_{{a\neq b}=1}^L\Psi_\mu^{ab}m_{\mu}^a(\bm\sigma) \Big).
\end{array}
\end{equation}
At $t=1$ we recover the original model, i.e.\ $\mathcal{A}_{N,K,L}^{(P,D)}(t=1)=\mathcal{A}_{N,K,L}^{(P,D)}(\bm{h},\beta,\lambda,\theta)$, while instead at $t=0$ the system reduces to a one-body (easily solvable) model: neurons decouple and interact only with an external field chosen to reproduce (at the level of the relevant low-order statistics) the internal field generated by the full network.
The relation between the two models is provided by the identity
\begin{equation}
    \label{eq:T_o_C}
    \mathcal{A}_{N,K,L}^{(P,D)}(t=1)= \mathcal{A}_{N,K,L}^{(P,D)}(t=0)+ \displaystyle\int_{0}^{1} dt\dfrac{\partial \mathcal{A}_{N,K,L}^{(P,D)}(s)}{\partial s}\Big\vert_{s=t}.
\end{equation}
We now consider the two contributions separately.
\par\medskip
\emph{One-body model.} The first contribution is easily solved. Starting from \eqref{eq:interpolating_Z} and setting $t=0$, we obtain
\begin{equation}
\begin{array}{lll}
       &\mathcal{Z}_{N,K,L}^{(P,D)}(t=0) = \SOMMA{\bm\sigma\in \Omega}{} \exp\Big( N\SOMMA{\mu=1}{L}\SOMMA{a=1}{L}\psi_\mu^a m_{\mu}^a(\bm\sigma)-N\sum\limits_{\mu=1}^{L} \sum\limits_{{a\neq b}=1}^L\Psi_\mu^{ab}m_{\mu}^a(\bm\sigma)+\beta \theta\sum\limits_{a=1}^{L}\sum\limits_{i=1}^{N}h_i^a \sigma_i^a\Big).
\end{array}
\end{equation}
Then, by \eqref{eq:interpol_pressure}:
\begin{equation}
\begin{array}{lll}
       &\mathcal{A}^{(P,D)}_{N,K,L}(t=0) =  \SOMMA{a=1}{L}\mathbb E_{\bm\xi} \ln\SOMMA{\bm\sigma^a}{} \exp\Big[ \big(\SOMMA{\mu=1}{L}\psi_\mu^a \xi^\mu-\sum\limits_{\mu=1}^{L}  \sum\limits_{b\neq a}^L\Psi_\mu^{ab}\xi^\mu+\beta \theta h^a \big)\sigma^a\Big].
\end{array}
\end{equation}
Summing over $\{\bm\sigma^{(a)}\}$ and using \eqref{eq:interpol_cons}, one gets
\begin{equation}
\label{eq:one-body}
\begin{array}{lll}
       &\mathcal{A}^{(P,D)}_{N,K,L}(t=0) =  \SOMMA{a=1}{L}\mathbb{E}_{\bm\xi}\log 2 \cosh\Big(
       \beta\SOMMA{\mu=1}{L}\big( \dfrac{P}{2}(\bar{m}^a_\mu)^{P-1}  - \sum\limits_{b\neq a}^L\lambda\dfrac{D}{2}(\bar{m}^a_\mu)^{D/2-1}(\bar{m}^b_\mu)^{D/2}\big)\xi^\mu + \beta \theta h^a\Big).
\end{array}
\end{equation}
\par\medskip
\emph{Remainder.}
We denote with $\omega_t(\cdot)$ the Boltzmann-Gibbs average w.r.t. the measure associated with the interpolating partition function in Eq. \eqref{eq:interpolating_Z}, and with $\mathbb E_{\bm\xi}\omega_t(\cdot)\equiv \langle\cdot\rangle_t$ its quenched expectation. Given a generic order parameter $x(\bb\sigma)$ we denote the thermodynamic limit of the quenched expectation as $\lim_{N\to\infty}\langle x(\bm\sigma)\rangle_t=\bar{ x}(t)$. With this definition, the $t$-derivative of $\mathcal{A}^{(P,D)}_{N,K,L}$ is
\begin{equation}
\label{eq:A_dot}
\begin{array}{lll}
      \dfrac{\partial\mathcal{A}_{N,K,L}}{\partial t} &=  \dfrac{\beta }{2} \SOMMA{a=1}{L}\SOMMA{\mu=1}{L}\l (m_{\mu}^a(\bm\sigma))^P\r_t -\SOMMA{\mu=1}{L}\SOMMA{a=1}{L}\psi_\mu^a \l m_{\mu}^a(\bm\sigma)\r_t
       \\\\
       &- \dfrac{\beta  \lambda}{2}\sum\limits_{\mu=1}^{L} \sum\limits_{{a\neq b=1}}^L  \l (m_{\mu}^a(\bm\sigma)m_{\mu}^b(\bm\sigma))^{D/2} \r_t + \sum\limits_{\mu=1}^{L} \sum\limits_{{a\neq b=1}}^L\Psi_\mu^{ab} \l m_{\mu}^a(\bm\sigma) \r_t
\\\\
    &=\dfrac{\beta }{2} \SOMMA{a=1}{L}\SOMMA{\mu=1}{L}\l (m_{\mu}^a(\bm\sigma))^P-\dfrac{2\psi_\mu^a}{\beta}  m_{\mu}^a(\bm\sigma) \r_t - \dfrac{\beta  \lambda}{2}\SOMMA{\mu=1}{L} \SOMMA{\underset{b\neq a}{a,b=1}}{L}  \l (m_{\mu}^a(\bm\sigma)m_{\mu}^b(\bm\sigma))^{D/2}  -\dfrac{2\Psi^{ab}_\mu}{\beta\lambda}   m_{\mu}^a(\bm\sigma) \r_t.
\end{array}
\end{equation}
In the low-storage regime, the order parameters $m_\mu ^a$ do not fluctuate w.r.t. their thermodynamic value $\bar m_\mu^a$, namely as self-averaging of the Mattis magnetizations:\footnote{Since the model operates in its low-load limit, self-averaging is expected to hold, and consequently the RS solution is exact in this setting.}
\begin{equation}
\lim_{N\to\infty}\langle ( m_\mu^a-\bar m_\mu^a)^2\rangle = 0,
\qquad \forall\,a,\mu.
\end{equation}
In this setting, we can fix the interpolating constants as
\begin{equation}
\label{eq:interpol_cons}
    \begin{array}{lll}
         \psi_\mu^a= \beta \dfrac{P}{2}(\bar{m}^a_\mu)^{P-1},&& \quad\Psi_\mu^{ab}= \lambda\beta\dfrac{D}{2}(\bar{m}^a_\mu)^{D/2-1}(\bar{m}^b_\mu)^{D/2},
    \end{array}
\end{equation}
so that \eqref{eq:A_dot} becomes (after neglecting the fluctuations of the magnetizations)
\begin{align}
\label{eq:streaming_apx}
    \dot{\mathcal{A}}_{N,K,L}= -\dfrac{\beta}{2}(P-1)\SOMMA{\mu=1}{L}\SOMMA{a=1}{L}(\bar{m}^a_\mu)^P +\dfrac{\beta\lambda(D-1)}{2}\SOMMA{\mu=1}{L}\SOMMA{\underset{b\neq a}{a,b=1}}{L}(\bar{m}^a_\mu \bar{m}^b_\mu)^{D/2}.
\end{align}

\par
\begin{figure}[h]
    \centering
    \includegraphics[width=0.48\linewidth]{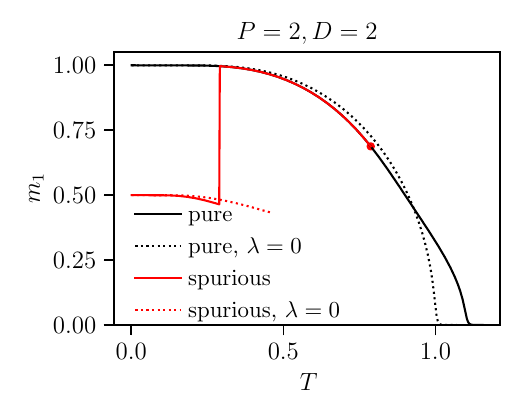}
    \includegraphics[width=0.48\linewidth]{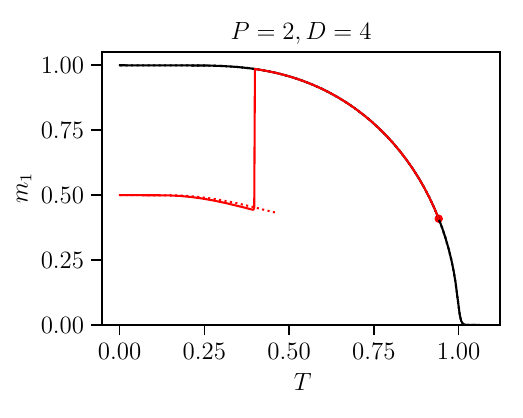}
    \includegraphics[width=0.48\linewidth]{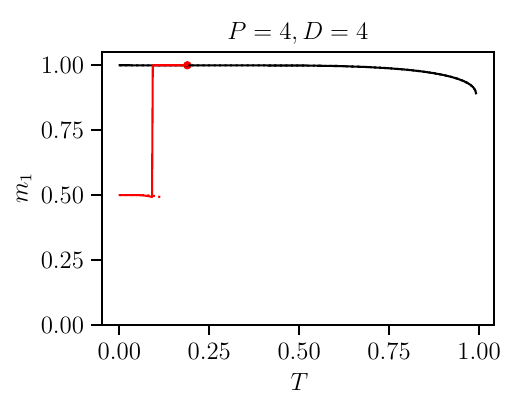}
    \includegraphics[width=0.48\linewidth]{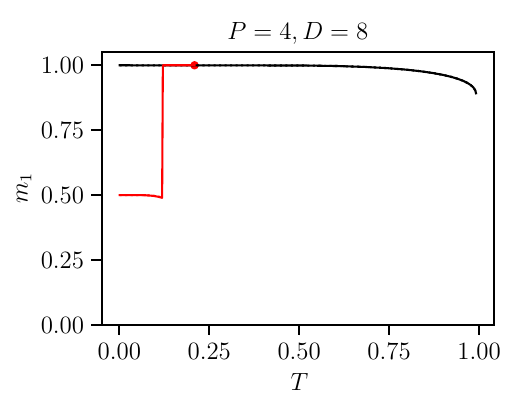}
    \caption{The figure shows the solution of the self-consistency equations for $\lambda=0.1$, $\theta=0$ and $L=3$ modules. The plots refer to $P=D=2$ (upper left), $P=2$, $D=4$ (upper right), $P=D=4$ (lower left) and $P=4$ and $D=8$ (lower right). In all cases, the black solid curves refer to the retrieval setting, namely the starting condition of fixed-point iteration is the retrieval setting (diagonal matrix for the magnetizations), while the red solid lines are derived starting from the $M=3$ spurious state $\operatorname{sign}(\bb\xi^1+\bb\xi^2+\bb\xi^3)$. The red dots represent the thermal noise level at which self-consistency equations prepared in the spurious state ceases to converge to the disentanglement scenario. Dashed curves refers to the $\lambda=0$ setting ($L$ independent copies of the dense Hopfield model) as a comparison baseline. In particular, the curve referring to spurious states is truncated up to the thermal level above which it becomes unstable.}
    \label{fig:sc_equations_solution}
\end{figure}

\medskip
\emph{Statistical pressure and self-consistency equations.}
Plugging \eqref{eq:streaming_apx} and \eqref{eq:one-body} in \eqref{eq:T_o_C}, we obtain the RS expression of the statistical pressure in the thermodynamic limit:
\begin{equation}
\label{eq:final_stat_pressure_RS}
\begin{array}{lll}
         \mathcal{A}_{N,K,L}^{(P,D)}(\bm{h},\beta,\lambda,\theta) &=& \SOMMA{a=1}{L}\mathbb{E}_{\bm\xi}\log 2 \cosh\Bigg[\beta\SOMMA{\mu=1}{L}\Big( \dfrac{P}{2}(\bar{m}^a_\mu)^{P-1}  - \sum\limits_{b\neq a}^L\lambda\dfrac{D}{2}(\bar{m}^a_\mu)^{D/2-1}(\bar{m}^b_\mu)^{D/2}\Big)\xi^\mu + \beta \theta h^a\Bigg]
         \\\\
         &-&\dfrac{\beta}{2}(P-1)\SOMMA{\mu=1}{L}\SOMMA{a=1}{L}\Big((
         \bar{m}^a_\mu)^P -\lambda\dfrac{D-1}{P-1}\SOMMA{b\neq a}{L}(\bar{m}^a_\mu \bar{m}^b_\mu\big)^{D/2}\Big).
\end{array}
\end{equation}
Extremizing the free energy \eqref{eq:final_stat_pressure_RS} with respect to $\bar{m}_{\nu}^{(a)}$ yields the self-consistency equations
\begin{equation}
\label{eq:one-body1}
\begin{array}{lll}
       &\bar{m}^a_\nu =\mathbb{E}_{\bm\xi}\xi^\nu\tanh\Big[\beta\SOMMA{\mu=1}{L}\big( \dfrac{P}{2}(\bar{m}^a_\mu)^{P-1}  - \sum\limits_{b\neq a}^L\lambda\dfrac{D}{2}(\bar{m}^a_\mu)^{D/2-1}(\bar{m}^b_\mu)^{D/2}\big)\xi^\mu  +\beta \theta h^a\Big],
\end{array}
\end{equation}
for $a=1,\dots,L$ and $\nu=1,\dots,K$. Importantly, the self-consistency equations do not depend on $\alpha$ as expected, as the dense model in the extensive scaling $K=\alpha N$ operates in its low-load regime, with globally magnetized equilibrium states extremizing the free energy regardless of the number of stored patterns. A summary of solutions of the self-consistency equations for $L=3$, $\lambda=0.1$ and $\theta=0$ as a function of $T$ and for different values of $P$ and $D$ is reported in Fig. \ref{fig:sc_equations_solution}. 

These solutions are determined as follows:
\begin{itemize}
    \item \emph{Disentangling solution.} In this case, each module specializes on retrieving a different pattern. By permutation symmetry of subnetworks, the desired solution has the form
    $$
    \bar {\bb m}^{dis}=\begin{pmatrix}
\bar m &0 &0\\
0 &\bar m &0\\
0 &0 &\bar m
\end{pmatrix}.
    $$
    \item \emph{Spurious states.} In this case, all modules relax to the same spurious state. Focusing on the $M=3$ scenario (leading to the most stable spurious state in the Hopfield model), the structure of the solution is
    $$
    \bar {\bb m}^{mix}=\begin{pmatrix}
\bar m &\bar m &\bar m\\
\bar m &\bar m &\bar m\\
\bar m &\bar m &\bar m
\end{pmatrix}.
    $$
\end{itemize}
Using these expression, we reduce the $L\times L$ set of self-consistency equations to a single implicit equation, whose solutions are respectively the black (referring to the disentangling solution) and red (spurious mixtures) curves. The dashed lines correspond to the $\lambda =0$ case (i.e. the pure dense Hopfield setting). As expected, switching on anti-imitative interactions generally destabilizes spurious states. The vertical jump signals the $\lambda^*$ value at which the spurious solution ceases to be stable, and small perturbations in the initial conditions results in the convergence on the disentangling solution. The phase diagrams in Fig. \ref{fig:TAM-phase-diag} are then realized by collecting the values $\lambda^*(\beta)$ yielding the vertical jumps and realizing the boundary for the spurious-to-disentangling transition (gray-to-yellow phases in those plots). The second boundary (disentangling-to-ergodic transition) is instead realized by estimating $\tilde \lambda(\beta)$ above which no disentangling solution with $\bar m_1>0$ is observed. Remarkably, the nature of this transition is always ruled by the auto-associative interaction order $P$: for the usual pairwise setting, the transition is critical (second-order), while it is discontinuous for $P>2$ (first-order).

\section{The effect of the encoding temperature}
\label{app:corr_input}
In this Appendix, we deepen the role of the encoding temperature $T_{enc}$ used to introduce additional noise in the spurious states $\bb\sigma_{\mu\nu \ell}$ used in the cryptographic Proof-of-Concept, see Sec. \ref{sec:crypto}. In particular, we compute the expected correlation $\mathbb E [\sigma_{\bm z} \xi^\mu]$ between the noisy variable  
$$\sigma_{\bm z}=\operatorname{sign}(\sum_{\ell=1}^K z_\ell \xi^\ell + T_{enc}\tilde{u}),$$
with $z_\ell \in \{0,1\}$, ${\xi}^\mu\sim_{i.i.d.} Rad$, $\mu=1,\dots,K$, and $\tilde u \sim\mathcal U[-1,+1]$, as a function of $T_{enc}\ge0$. This mimics the site-wise correlation between the noisy input mixture and the stored patterns. 
%
Using $\xi^\mu\in\{-1,+1\}$ gives
\[
\mathbb{E}[
\operatorname{sign}(\sum_{\ell=1}^K z_\ell \xi^\ell + T_{enc}\tilde{u})\xi^\mu
]
=
\mathbb{E}[
\operatorname{sign}(\xi^\mu\sum_{\ell=1}^K z_\ell \xi^\ell + T_{enc}\tilde{u}\,\xi^\mu)
],
\]
where $\mathbb E$ stands for the expectation w.r.t. the joint law of $\xi$ and the noise $\tilde u$. Since $\tilde{u}$ is symmetric, the random variable $\tilde{u}\xi^\mu$ has the same law as $\tilde{u}$, then
\[
\mathbb{E}\!\left[\sigma_{\bm z} \xi^\mu\right]
=
\mathbb{E}[
\operatorname{sign}(\xi^\mu\sum_{\ell=1}^K z_\ell \xi^\ell + T_{enc}\tilde{u})
].
\]
Without loss of generality, we can assume $z_\ell =1$ for $\ell \le L$\footnote{Since the cryptographic protocol is designed to have the equal number of modules and number of components in the spurious states used for transmission, here we directly specialize to $L=M$.} and $0$ otherwise. 
If $\mu\notin\{1,\dots,L\}$, the expectation vanishes by symmetry. For $\mu\leq L$
\[
\xi^\mu\sum_{\ell=1}^L \xi^\ell
=
1+\sum_{\substack{\ell=1\\\ell\neq\mu}}^L \xi^\mu\xi^\ell
=
1+\sum_{\ell=1}^{L-1}\rho^\ell,
\]
where the $\rho^\ell$ are i.i.d. Rademacher random variables. Therefore
\[
\mathbb{E}\!\left[\sigma_{\bm z} \xi^\mu\right]
=
\mathbb{E}[
\operatorname{sign}(1+\sum_{\ell=1}^{L-1}\rho^\ell+T_{enc}\tilde{u})
].
\]
From this representation, the behavior in the different noise regimes becomes transparent. We now focus on odd $L$ (as spurious states with odd number of components are stable in the Hopfield setting); then, $L-1$ is even, and $\sum_{\ell=1}^{L-1}\rho^\ell$ is the sum of an even number of $\pm1$ variables. In what follows, it is convenient to define $x=1+\sum_{\ell=1}^{L-1} \rho^\ell$.
\par\medskip
We first consider the computation at $T_{enc}=0$:
\[
\mathbb{E}[\sigma_{\bm z} \xi^\mu]
=
\mathbb{E}[
\operatorname{sign}(x)
].
\]
Since $\sum_{\ell=1}^{L-1}\rho^\ell$ can only take even values and it is a symmetric random variables, the previous quantity reduces to the probability that $\sum_{\ell=1}^{L-1}\rho^\ell=0$, namely
\[
\mathbb{E}\!\left[\sigma_{\bm z} \xi^\mu\right]
=
\frac{1}{2^{L-1}}
\binom{L-1}{\frac{L-1}{2}}
\doteq \mathcal C_L.
\]
We now consider $T_{enc}>0$.  A direct average over the uniform noise gives
\[
\mathbb{E}_u\!\left[\operatorname{sign}(x+uT_{enc})\right]
=
\begin{cases}
\operatorname{sign}(x)& \vert x\vert > T_{enc}\\
 x/{T_{enc}} & |x|\le T_{enc}
\end{cases},
\]
or equivalently
$$
\mathbb{E}_u\!\left[\operatorname{sign}(x+uT_{enc})\right]-\operatorname{sign}(x) = \operatorname{sign}(x)\Big(\frac{\vert x\vert}{T_{enc}}-1\Big)1\{|x|\le T_{enc}\}.
$$
Taking the expectation w.r.t. the realization of the patterns one has
$$
\mathbb E_{\rho}\mathbb{E}_u\!\left[\operatorname{sign}(x+uT_{enc})\right]-\mathcal C_L =\mathbb E_{\rho} \operatorname{sign}(x)\Big(\frac{\vert x\vert}{T_{enc}}-1\Big)1\{|x|\le T_{enc}\},
$$
since $\mathbb E_{\rho} \operatorname{sign}(x)= \mathcal C_L$ by definition. Since the random walk $S=\sum_{\ell=1}^{L-1}\rho^\ell$ can take even values between $-L+1$ and $L-1$ with spacings 2, we can parametrize positive values of $x= 2r+1$ (corresponding to $S=2r$), while for the negative values $x=-(2r+1)$ (corresponding to $S=-(2r+2)$). Calling $q$ the number of contributions satisfying $ x  \le T_{enc}$ consistently with the natural bound $ x \le L $, we have
\begin{equation*}
    \begin{split}
       \mathbb E_{\rho} \operatorname{sign}(x)\Big(\frac{\vert x\vert}{T_{enc}}-1\Big)1\{|x|\le T_{enc}\} &= \sum_{r=0}^{q-1} \Big(\frac{2r+1}{T_{enc}}-1\Big) [P(S=2r)-P(S=-2r-2)].
    \end{split}
\end{equation*}
Since the random walk $S$ is symmetric (as $\rho^\ell$ is Rademacher random variable), we have $P(S=-2r-2)=P(S=2r+2)$. But
\begin{equation*}
    \begin{split}
    P(S=2r)-P(S=2r+2)&=\frac{1}{2^{L-1}}\Big[\binom{L-1}{\frac12(L-1+2r)}-\binom{L-1}{\frac12(L+1+2r)}\Big]=\\
    &=\frac{1}{2^{L-1}}\binom{L-1}{\frac12(L-1+2r)}\frac{2(2r+1)}{L+1+2r},
    \end{split}
\end{equation*}
where the last equality follows from binomial coefficient identities. Putting everything together leads to
\[
\mathbb E\!\left[\sigma_{\bm z} \xi^\mu\right]
=
\mathcal C_L-
\sum_{r=0}^{q-1}
\left(1-\frac{2r+1}{T_{enc}}\right)
\frac{2(2r+1)}{L+2r+1}\,
\frac{1}{2^{L-1}}
\binom{L-1}{\frac{L+2r-1}{2}}.
\]
holding for odd $L$ and with
\[
q=\min\Big(\left\lfloor \frac{T_{enc}+1}{2}\right\rfloor,\frac{L+1}{2}\Big).
\]

\begin{figure}[t]
    \centering
    \includegraphics[width=8cm]{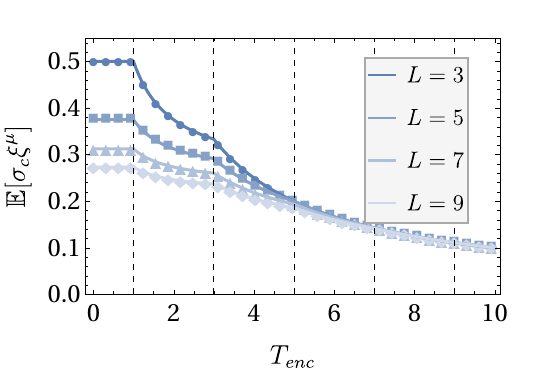}
    \caption{
    Expected correlation $\mathbb{E}\!\left[\sigma_c \xi^\mu\right]$ as a function of the noise amplitude $T_{enc}$, for different values of $L$. Dots denote numerical simulations, while solid lines represent the theoretical prediction. The dashed colored lines are set to $T_{enc}=1,3,5,7$ respectively.
    }
    \label{fig:noise_correlation}
\end{figure}

This expression correctly reproduces both limiting behaviors discussed above.
The resulting theoretical prediction is compared with numerical simulations in Fig.~\ref{fig:noise_correlation}, where one can clearly observe the crossover from the low-noise plateau $\mathcal C_L$ to the asymptotic decay $1/T_{enc}$ at large noise.

\section{Effectiveness of the masking procedure}\label{app:masking}

\begin{figure}[h!]
    \centering
    \includegraphics[width=15cm]{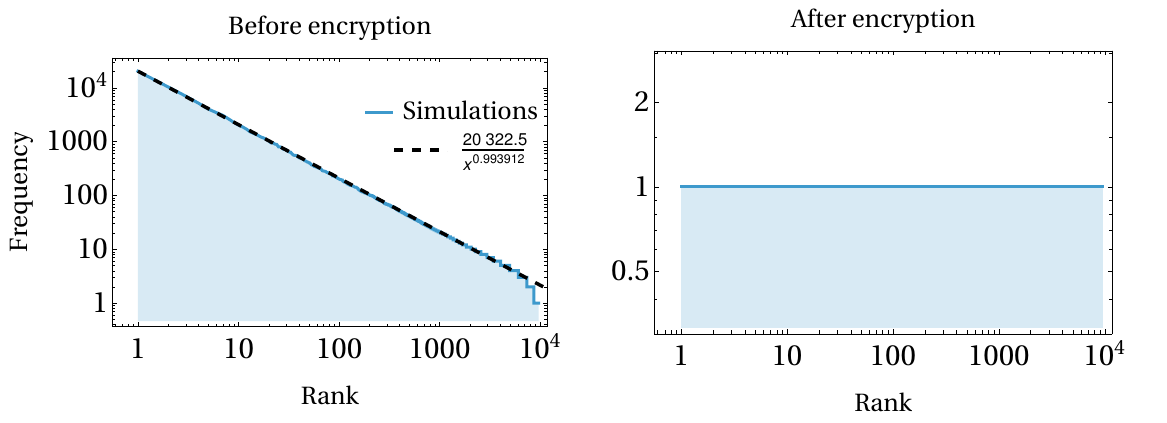}
    \caption{Statistical flattening of the rank--frequency distribution induced by the encryption scheme. Left panel: empirical token frequency versus rank in the plaintext corpus, showing the expected Zipf-like heavy-tailed behavior. Right panel: empirical frequency versus rank for the encrypted representations of the same corpus. Since repeated occurrences of the same word are mapped to different ciphertext realizations by the stochastic encoding procedure and the additional pseudorandom masking, the rank information is destroyed. The ciphertext distribution becomes essentially flat, with frequencies close to one for all observed encrypted tokens.}
    \label{fig:crypto_zipf_flattening}
\end{figure}
To better evaluate the performance of the masking scheme used in Sec. \ref{sec:crypto}, we designed a robust set of experiments. Sentences are generated by repeatedly sampling tokens from a subset of the dictionary consisting in $341$ words (not including punctuation marks) according to a predefined unigram probability vector $\bm{p}$. This ensures that high-rank function words are drawn much more frequently than low-rank content words, closely matching the heavy-tailed distribution typical of the English language. The vector $\bm{p}$ is constructed by assigning each word $w_i$ a weight derived from its usage frequency, estimated via the \texttt{wordfreq} library \cite{robyn_speer_2022_7199437}. Specifically, we query \texttt{zipf\_frequency}$(w_i,\text{``en''})$ to obtain a Zipf-scale score $z_i$, convert it to a linear weight $\tilde{w}_i = 10^{z_i}$, and normalize across the vocabulary to compute the probabilities $p_i=\tilde{w}_i/\sum_j \tilde{w}_j$. To mimic natural text segmentation, punctuation marks are inserted stochastically at a controlled rate, while lexical tokens are sampled independently from $\bm{p}$. To confirm that our sampling reflects the expected statistics, we generated $10^4$ random sentences of fixed length ($150$ tokens), computed the empirical word counts, and plotted the frequency versus rank on log--log axes. As shown in the left panel of Fig.~\ref{fig:crypto_zipf_flattening}, the simulated frequencies closely follow a Zipf-like power law, aligning perfectly with the overlaid fit. Although this synthetic dataset exhibits a strong, easily identifiable statistical distribution -- which could potentially allow an external observer to infer the transmitted messages -- our encryption mechanism successfully mitigates this vulnerability. Consequently, even if the same plaintext word is transmitted multiple times, the observable object is never the clean spurious state, but rather its masked counterpart. Since the legitimate receiver removes this mask before proceeding with decryption, standard decoding remains entirely unaffected. Conversely, an external attacker only has access to the masked ciphertext. This introduces a second layer of security, complementing the fact that the hidden patterns are never individually transmitted. To quantitatively assess the statistical obfuscation induced by our scheme, we performed two complementary analyses. The first focuses on the rank-frequency statistics of the plaintext versus the ciphertext. While natural language heavily relies on a Zipf-like distribution -- a structure deliberately replicated by our plaintext generator -- this property is entirely eradicated after encryption. Since the encoding stage is stochastic and further masked, repeated occurrences of the same lexical token do not map to a unique observable ciphertext. As a result, the empirical frequency distribution of the encrypted tokens becomes almost completely flat; each observed encrypted realization appears only once, or with negligible repetition. By destroying the lexical rank information of the original message, the ciphertext prevents an attacker from building a frequency-based empirical dictionary. This statistical flattening is clearly illustrated in the right panel of Fig.~\ref{fig:crypto_zipf_flattening}. Furthermore, the fact that highly frequent tokens rarely produce the same transmitted representation means an external observer cannot reliably accumulate occurrence statistics over time. 
\par\medskip

\begin{figure}[h!]
    \centering
    \includegraphics[width=15cm]{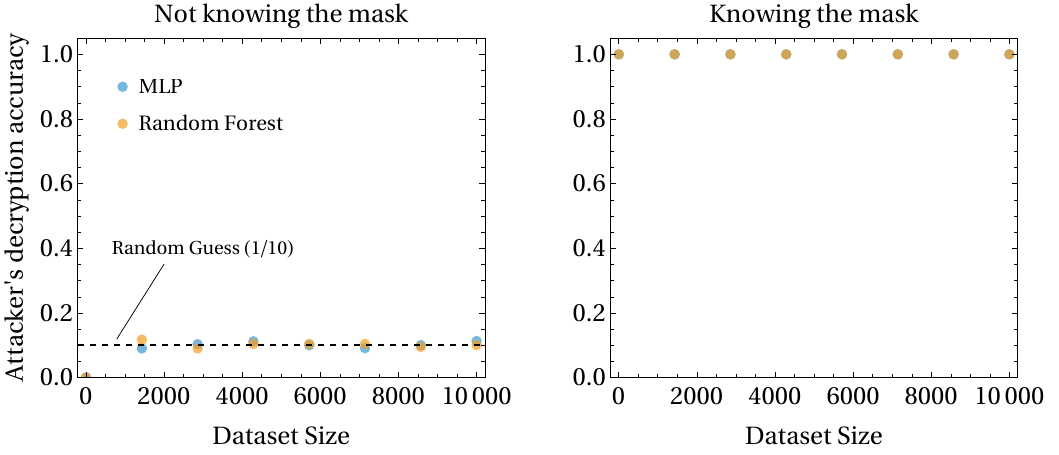}
    \caption{Attacker's decryption accuracy in a supervised learning attack, shown as a function of the dataset size, for a dictionary of $10$ classes. We compare a multilayer perceptron (MLP) and a random forest. Left panel: if the masking key $\bm{\eta}$ is unknown, even an extremely large collected dataset is essentially useless to the attacker, since the accuracy stays at the chance level ($1/10$). Right panel: if $\bm{\eta}$ is known, the classification problem becomes trivial and both models reach nearly perfect accuracy. Hence, the pseudorandom mask $\bm{\eta}$ is what effectively suppresses exploitable regularities in the ciphertext.}
    \label{fig:AI_attack}
\end{figure}

The protective role of the pseudorandom mask can also be evaluated against data-driven inference attacks. To this end, we simulate a supervised AI-based attack in which an external observer collects encrypted tokens generated from a reduced dictionary and attempts to infer the corresponding plaintext label directly from the observed ciphertext. More precisely, we consider a dictionary composed of only $10$ words and assume that the attacker has collected pairs of ciphertexts and and the corresponding plaintexts, thereby constructing a dataset of tuples of the form $(\text{ciphertext}, \text{plaintext})$. The attacker then trains an AI model that takes the ciphertext as input and predicts which of the $10$ dictionary words generated it. We repeat this experiment for different training-dataset sizes and evaluate the resulting model on an independent test set of $10^4$ ciphertext-plaintext tuples. We trained both a multilayer perceptron (MLP) and a random-forest classifier on datasets of increasing size.\footnote{For MLP we use \texttt{MLPClassifier$()$} from \texttt{sklearn.neural$\_$network} with only on hidden layer of size 256, and Relu activation function; for Random Forest Classifier we use \texttt{RandomForestClassifier()} from \texttt{sklearn.ensemble} with 100 estimators.} As shown in Fig.~\ref{fig:AI_attack}, when the attacker lacks knowledge of the masking key $\bm{\eta}$, prediction accuracy remains essentially at the random-guessing baseline, even for extensively large training sets. Conversely, once the mask is revealed, the classification task becomes trivial, and both models achieve nearly perfect accuracy. This provides direct numerical evidence that the shared pseudorandom mask $\bm{\eta}$ is a central security layer, suppressing exploitable regularities in the ciphertext and strongly hindering empirical dictionary reconstruction via AI-based attacks.

Together, these results demonstrate that the proposed associative-memory-based encryption scheme achieves robust statistical masking. While this empirical validation should not be strictly interpreted as a formal proof of information-theoretic or computational security, it provides clear numerical evidence that the approach effectively prevents basic frequency analysis and statistical attacks.



\section{Improving in repetitions}
\label{app:repetitions}

\begin{figure}[t!]
    \centering
    \includegraphics[width=\linewidth]{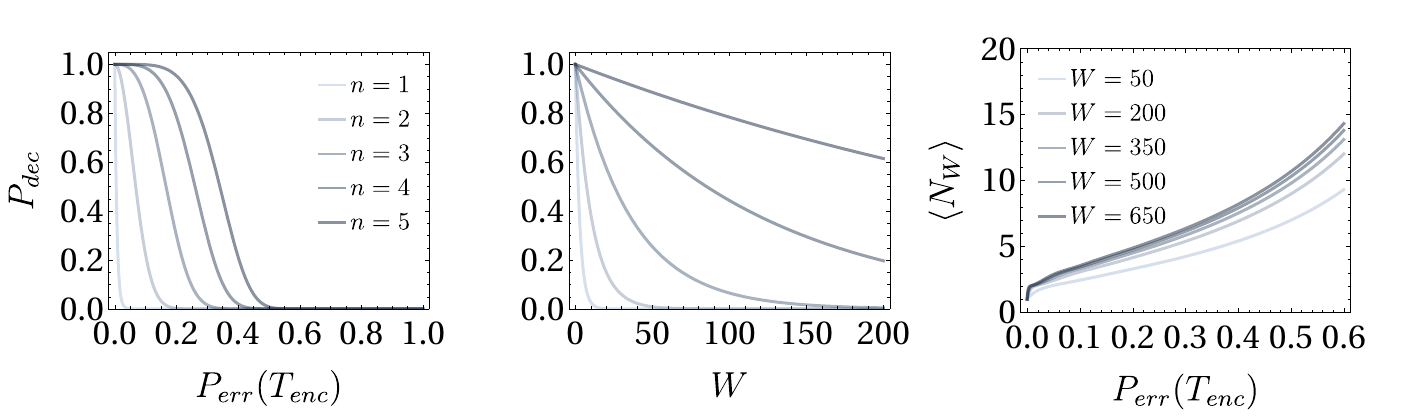}
    \caption{
    Left: decoding probability \(P_{\mathrm{dec}}\) as a function of the word-reconstruction error probability \(P_{\mathrm{err}}(T_{\mathrm{enc}})\), for different numbers of repeated full-message transmissions for fixed value of \(W=150\). Increasing the number of repetitions shifts the decoding threshold towards larger values of \(P_{\mathrm{err}}(T_{\mathrm{enc}})\), indicating an enhanced robustness against transmission errors.
    Center: decoding probability \(P_{\mathrm{dec}}\) as a function of the message length \(W\), for different numbers of repeated transmissions for fixed value of \(P_{\mathrm{err}}(T_{\mathrm{enc}})=0.3\).
    Right: average number of full message transmissions, \(\langle N_W\rangle\), required for complete reconstruction of a message of length \(W\), as a function of the encoding temperature \(T_{\mathrm{enc}}\). The quantity \(\langle N_W\rangle\) is obtained from Eq.~\eqref{eq:mean_transmissions} using the measured word-reconstruction error probability \(P_{\mathrm{err}}(T_{\mathrm{enc}})\).
    }
    \label{fig:prob_error}
\end{figure}

Due to the random nature of errors in the system and the stochasticity of the decoding neural network, repeated transmissions improve overall decoding reliability as $n$ increases. Let us consider a message composed of \(W\) words. We denote with \(P_{\mathrm{err}}(T_{\mathrm{enc}})\) the word-reconstruction error probability as a function of the encoding temperature \(T_{\mathrm{enc}}\) (see Fig.~\ref{fig:crypto_performance}). Transmission errors are assumed to be independent from word to word and from one full message transmission to the next. Successive transmissions are performed by sending the entire message again, word by word. For a single transmission of the message, let \(X_i\) be the indicator variable associated with the \(i\)-th word,
\[
X_i =
\begin{cases}
1, & \text{if the \(i\)-th word is reconstructed incorrectly},\\
0, & \text{otherwise}.
\end{cases}
\]
Since each word is corrupted independently with probability \(P_{\mathrm{err}}(T_{\mathrm{enc}})\), $X_i$ is a Bernoulli random variable (with 1 occurrence meaning an erroneous decoding); then 
\[
\langle X_i \rangle = P_{\mathrm{err}}(T_{\mathrm{enc}}).
\]
The total number of errors in one full transmission is therefore $X = \sum_{i=1}^{W} X_i$, and the average number of transmission errors in a single message is $\langle X \rangle= W\,P_{\mathrm{err}}(T_{\mathrm{enc}}).$ This implies that, at fixed message length \(W\), the average number of corrupted words directly follows the behavior of \(P_{\mathrm{err}}(T_{\mathrm{enc}})\).
\par\medskip
We now consider the number of full transmissions required for complete reconstruction of the message. The receiver is allowed to compare several independently transmitted copies of the same message and reconstruct each word as soon as that word has been received correctly at least once. The message is completely reconstructed when every one of the \(W\) words has appeared correctly at least once among the received copies. Let \(N_W\) denote the number of full message transmissions required for complete reconstruction. For a given word, the probability that it is still incorrect after \(n\) independent transmissions is
\[
\bigl(P_{\mathrm{err}}(T_{\mathrm{enc}})\bigr)^n,
\]
since it must be corrupted in every transmission. Therefore, the probability that this word has been received correctly at least once after \(n\) transmissions is
\[
1-\bigl(P_{\mathrm{err}}(T_{\mathrm{enc}})\bigr)^n.
\]
Because the \(W\) words are independent, the probability that all \(W\) words have been observed correctly at least once after \(n\) transmissions is
\begin{equation}
P_{dec}=\mathcal{P}(N_W \le n)
=
[1-(P_{\mathrm{err}}(T_{\mathrm{enc}}))^n]^W,
\end{equation}
or equivalently,
\begin{equation}
    \mathcal{P}(N_W > n) = 1-P_{dec}.
\end{equation}
This dependence is illustrated in the upper row of Fig.~\ref{fig:prob_error}, where we report \(P_{dec}\) for different values of the number of transmissions $n$ as a function of the encoding temperature and of the message length \(W\). In the ideal setting of infinitely many transmissions of the same message, since \(N_W\) is a positive integer-valued random variable, its expectation can be written as\footnote{In the finite transmission attempts, the expectation would be $\langle N_w\rangle = \sum_{n=0}^{n_{max}} \mathcal P (N_W>n)$, with $n_{max}$ being the total number of transimissions of the same message.}
\begin{equation}
\langle N_W \rangle
=
\sum_{n=0}^{\infty}\mathcal{P}(N_W>n)
=
\sum_{n=0}^{\infty}
(
1-P_{dec}
).
\label{eq:mean_transmissions}
\end{equation}
This is the exact expression for the average number of complete message transmissions required for full reconstruction, at fixed sentence length \(W\) and word-reconstruction error probability \(P_{\mathrm{err}}(T_{\mathrm{enc}})\). The dependence of \(\langle N_W\rangle\) on the encoding temperature, as obtained from Eq.~\eqref{eq:mean_transmissions}, is shown in Fig.~\ref{fig:prob_error}. As expected, \(\langle N_W\rangle\) increases rapidly in the regime where \(P_{\mathrm{err}}(T_{\mathrm{enc}})\) becomes large, reflecting the increasingly large number of repeated transmissions needed to observe each word correctly at least once.

\end{document}